\documentclass[twocolumn,pre,reprint,showpacs,preprintnumbers,amsmath,amssymb,superscriptaddress]{revtex4-2}
\usepackage{graphicx}
\usepackage{dcolumn}
\usepackage{bm}
\usepackage{amsmath}
\usepackage{physics}
\usepackage{multirow}
\newcommand{\tens}[1]{%
	\mathbin{\mathop{\otimes}\limits_{#1}}%
}

\let\oldAA\AA
\renewcommand{\AA}{\text{\normalfont\oldAA}}

\usepackage{upgreek}
\usepackage{amsmath}
\usepackage{subfigure}
\usepackage{hyperref}
\usepackage[normalem]{ulem}
\hypersetup{
    colorlinks,
    citecolor=blue,
    filecolor=black,
    linkcolor=blue,
    urlcolor=blue
}
\usepackage{epstopdf}
 \usepackage{relsize}

\newcommand*{\rom}[1]{\expandafter\@slowromancap\romannumeral #1@}
\title{Report}

\setcitestyle{square}

\begin{document}


\title{A  non-local cryogenic thermometer  based on Coulomb-coupled systems }
\author{Sagnik Banerjee}
\thanks{These two authors contributed equally to this work. Any correspondence should be sent at aniket@ece.iitkgp.ac.in.}
\affiliation{%
	Department of Electronics and Telecommunication Engineering, \\Jadavpur University,  Jadavpur-700032, India\\
}%
\author{Aniket Singha}

\thanks{These two authors contributed equally to this work. Any correspondence should be sent at aniket@ece.iitkgp.ac.in.}


\affiliation{%
Department of Electronics and Electrical Communication Engineering, \\Indian Institute of Technology Kharagpur, Kharagpur-721302, India\\
}%






\begin{abstract}
 We investigate  a  quadruple quantum dot set-up that can be employed to sense temperature of an electrically isolated remote target reservoir. Such a set-up was  conceived earlier by  S\'anchez \emph{et. al.} as non-local thermodynamic-engine (\emph{New Journal of Physics, 19, 113040}), and relies on the electrostatic interaction between Coulomb-coupled quantum dots. The conjugation of Coulomb-coupling and energy-filtering results in an overall change in conductance with remote reservoir temperature.  \color{black} The performance of the thermometer is then theoretically investigated using density matrix formulation, and it is demonstrated that the quadruple quantum dot design ensures a superior temperature sensitivity and noise robustness compared to a simple thermometer consisting of two Coulomb-coupled quantum dots. In the end, we investigate the regime of operation and comment on the ground state configuration for optimal performance of the thermometer. The  set-up investigated in this paper can be employed to construct highly efficient non-local cryogenic thermometers.

\end{abstract}
\maketitle
\section{Introduction}
Electrical sensing of temperature and heat flow in nano-scale systems, particularly in the cryogenic regime,  has been one of the biggest challenges in the modern era. Engineering devices to couple heat with electrically measurable quantities has been extremely difficult in the domain of solid-state nano-technology. In the aspect of thermally controlled electrical transport in nano-scale systems; thermoelectric engines \cite{thermoelectric_transport_at_mesoscopic_level_1,aniket_nonlocalheat,thermoelectric_transport_at_mesoscopic_level_2,heatengine1,heatengine2,heatengine3,heatengine4,heatengine5,aniket,heatengine6,bd1,bd2,bd3,aniket_heat1,aniket_heat2,heatengine7,heatengine8,heatengine9,heatengine10,heatengine11,heatengine12,heatengine13,heatengine14,heatengine15,heatengine16,heatengine17}, refrigerators \cite{thermalrefrigerator1,thermalrefrigerator2,aniket_nonlocalref,aniket_cool1,aniket_cool2,thermalrefrigerator3,thermalrefrigerator4,thermalrefrigerator5,thermalrefrigerator6,bd4,thermalrefrigerator7,thermalrefrigerator8}, rectifiers \cite{thermalrectifier1,thermalrectifier2,thermalrectifier3,thermalrectifier4,thermalrectifier5,thermalrectifier6} and transistors \cite{thermaltransistors1,thermaltransistors2,thermaltransistors3,thermaltransistors4,thermaltransistors5,thermaltransistors6,thermaltransistors7,thermaltransistors8} have been proposed in the last decade. Recently, the provision towards \emph{non-local} thermal control of electrical transport has been proposed and realized experimentally \cite{3terminalheatengine1,3terminalheatengine2,3terminalheatengine3,3terminalheatengine4,3terminalheatengine5,3terminalheatengine6,3terminalheatengine7,3terminalheatengine8,3terminalheatengine9,3terminalrefrigerator1,3terminalrefrigerator2,3terminalrefrigerator3,3terminalrefrigerator4}. In this case, electrical variables between two terminals are manipulated via thermal action at a remote  terminal,  which acts as the heat source (sink) or the control reservoir. The remote terminal is spatially and electrically isolated from the path of current flow, thereby prohibiting any exchange of electrons with the current conduction track. \color{black} \\
\indent Thus, proposals towards non-local thermal control of electrical transport mainly include multi-terminal devices, where current/voltage between two terminals may be manipulated via heat energy stored at one or many non-local remote reservoirs \cite{3terminalheatengine1,3terminalheatengine2,3terminalheatengine3,3terminalheatengine4,3terminalheatengine5,3terminalheatengine6,3terminalheatengine7,3terminalheatengine8,3terminalheatengine9,3terminalrefrigerator1,3terminalrefrigerator2,3terminalrefrigerator3,3terminalrefrigerator4}. Non-local coupling between thermal and electrical quantities offer many distinct advantages over their local counterparts, which include (but are not limited to) the provision towards an independent manipulation of electrical and lattice thermal conductance, isolation of the remote target reservoir from Joule heat dissipation, etc. Proposals towards temperature-induced control of electronic flow in quantum dots \cite{sensor_qdot}, and quantum-point-contacts \cite{sensor_qpc}, due to stochastic thermal fluctuation in a capacitively coupled quantum dot, have already been put-up in literature. In addition, a lot of  effort has been geared towards theoretical and experimental realization of electrical thermometers \cite{thermometer1,thermometer2,thermometer3,thermometer4,milikelvinthermometer1,milikelvinthermometer2,milikelvinthermometer3,milikelvinthermometer4,milikelvinthermometer5,milikelvinthermometer6,milikelvinthermometer7,milikelvinthermometer8,milikelvinthermometer9,milikelvinthermometer10,milikelvinthermometer11}.  \\
\indent In this paper, we investigate a quadruple quantum dot set-up that can be used to sense temperature from a remote electrically isolated reservoir. The spatial separation of the target reservoir from the current flow track not only shields the reservoir from unnecessary Joule heating, but also offers the provision towards independent manipulation of lattice thermal conductance, such that good thermal isolation of the target reservoir can be achieved. Although electrical thermometers based on Coulomb-coupled quantum dots have already been put up in literature \cite{sensor_qdot}, the optimal operation of such thermometers demands a sharp step-like transition in the system-to-reservoir coupling, which is impossible to achieve in practical fabricated systems. Additionally, the sensitivity of such a system is dependent on the average current path temperature \cite{sensor_qdot}.   On the contrary, the design considered in this paper circumvents the demand for any change or transition in the system-to-reservoir coupling for optimal operation.  In addition, the current set-up offers a superior temperature sensitivity and noise robustness compared to the simple thermometer investigated in \cite{sensor_qdot}. Also, the thermometer sensitivity for sufficiently high bias voltage is independent of the average  current  path temperature, making such systems suitable for practical purposes.  \\
\indent This paper is organized as follows. In Sec. \ref{design}, we elaborate on the system under consideration and briefly describe the transport formulation employed to analyze its performance. The detailed derivations for the transport formulation are given in Appendix \ref{app_a}. In Sec. \ref{results}, we discuss the regime of operation and performance of the thermometer. This section also presents a brief discussion on the  performance of  the quadruple quantum dot thermometer with respect to  a simple thermometer based on two Coulomb-coupled quantum dots. Finally, we conclude the paper briefly in Sec. \ref{conclusion}.
 \begin{figure*}
 	\includegraphics[width=1\textwidth]{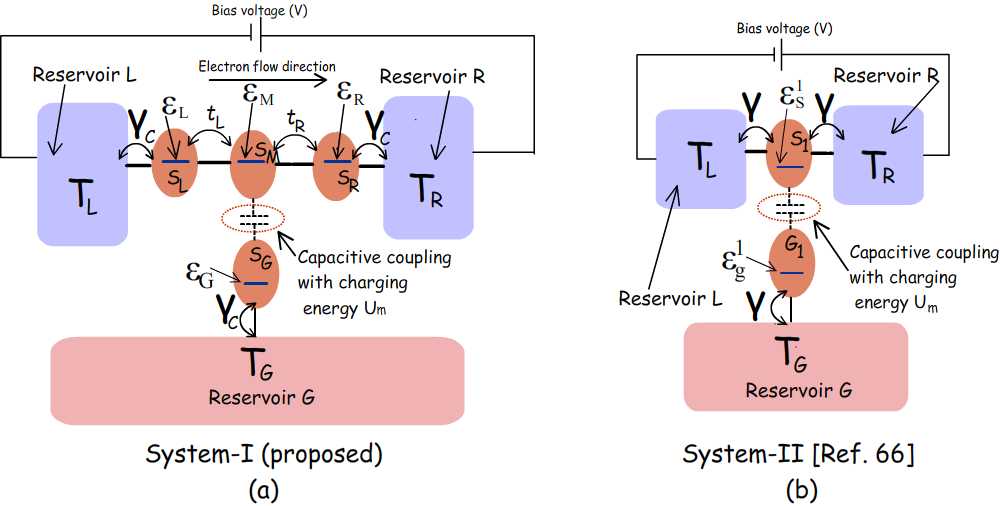}
 	\caption{ (a) Schematic diagram of the considered non-local cryogenic electrical thermometer based on Coulomb-coupled quantum dots. The entire system consists of four dots $S_L$, $S_M$, $S_R$, $S_G$. The dots $S_L$, $S_R$, and $S_G$ are electrically coupled to reservoirs \textit{L}, \textit{R}, and \textit{G} respectively. The dot $S_M$ is electrically coupled to both $S_L$ and $S_R$, while $S_M$ and $S_G$ are capacitively coupled to each other (with Coulomb-coupling energy $U_m$). The ground state energy levels of the three dots $S_L$, $S_M$ and $S_R$, denoted by    $\varepsilon_L$, $\varepsilon_M$ and $\varepsilon_R$ respectively, are aligned with each other for maximum conductance between $L$ and $R$. The parameter $\gamma_c$  denotes the reservoir-to-system tunnel coupling, while $t_L$ and $t_R$ denote the inter-dot tunnel coupling amplitudes.  This thermometer is based on the non-local thermodynamic engine set-up proposed by S\'anchez, \emph{et. al.} \cite{thermaltransistors4}. \color{black} We will designate this set-up as system-I (b)  Schematic of the recently proposed electrical thermometer based on Coulomb-coupled systems \cite{sensor_qdot}.   This thermometer set-up is based on a simpler thermodynamic engine proposed by S\'anchez \emph{et. al.} \cite{heatengine5} and consists of two Coulomb-coupled quantum dots $S_1$ and $G_1$ \color{black}. $S_1$ is electrically connected to the reservoirs $L$ and $R$ and provides the path for current flow. $G_1$ on the other hand, is electrically connected to the remote reservoir $G$ whose temperature is to be accessed. We will call this system as system-II}
 	\label{fig:Fig_1}
 \end{figure*}
  \section{Design and transport formulation}\label{design}
\begin{figure}[!htb]
 	\includegraphics[width=0.35\textwidth]{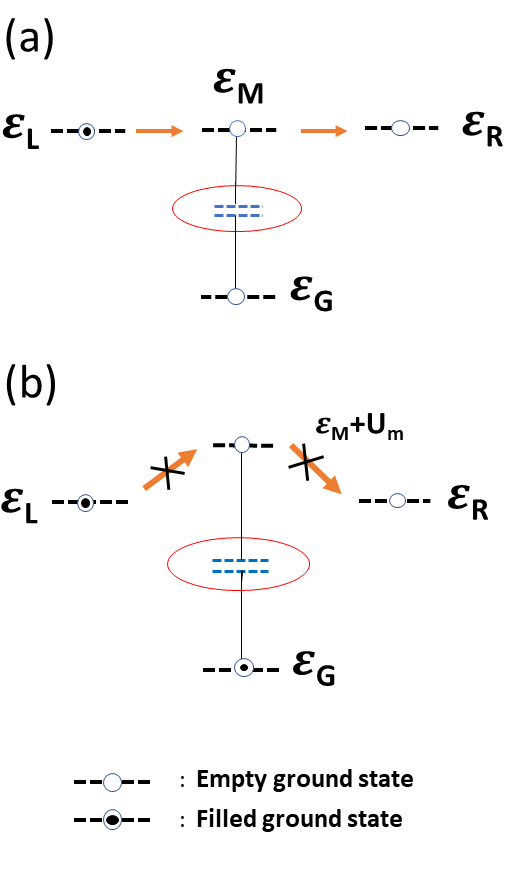}
 	\caption{Schematic diagram depicting the electronic transport process through  system-I. (a) Case-I: No electron is present in the ground state $\varepsilon_G$. In this case the ground states of the dots $S_L$, $S_M$ and $S_R$ are aligned with each other and electron can flow between the reservoirs $L$ and $R$ via the quantum dot ground states. 
 		(b) Case-II: An electron is present in the ground state $\varepsilon_G$. As dots $S_M$ and $S_G$ are capacitively coupled, in this case the Coulomb-blocked  ground state $\varepsilon_M+U_m$ is not aligned with the ground states of $S_L$ and $S_R$. This restricts the current flow through the system, since an electron can't easily tunnel into (or out of) the ground state of $S_M$. }
 	\label{fig:Fig_2}
 \end{figure}

\indent  The quadruple quantum dot thermometer, schematically demonstrated in Fig.~\ref{fig:Fig_1}(a) as system-I, is based on the non-local thermodynamic engine originally conceived by S\'anchez, \emph{et. al.} \cite{thermaltransistors4}.  It consists of four quantum dots $S_L$, $S_M$, $S_R$, and  $S_G$. \color{black} The dots $S_L$, $S_R$, and $S_G$ are electrically coupled to reservoirs \textit{L}, \textit{R}, and \textit{G} respectively, $G$ being the target reservoir whose temperature is to be accessed. The ground state energy levels of the four dots $S_L$, $S_M$, $S_R$ and  $S_G$ are denoted by  $\varepsilon_L$, $\varepsilon_M$, $\varepsilon_R$ and  $\varepsilon_G$ respectively, where  $\varepsilon_L$, $\varepsilon_M$, and $\varepsilon_R$ are aligned with each other. The dot $S_M$ is electrically coupled to both $S_L$ and $S_R$, while being capacitively coupled to  $S_G$.  The capacitive coupling permits exchange of electrostatic energy between the dots $S_M$ and $S_G$ while restricting any electronic flow between them, resulting in \textit{zero} net electronic current out of (into) the reservoir \textit{$G$} \cite{3terminalheatengine8}. \color{black}Coming to the provision towards the practical fabrication of such a system, the considerable advancement in solid-state nano-fabrication technology has led to the realization of three and four dot systems with and without Coulomb-coupling \cite{mqd1,mqd2,mqd3,mqd4,mqd5,mqd6}. In addition, two specific quantum dots that are far in space may be bridged together to achieve strong capacitive coupling, without affecting the electrostatic energy of adjacent quantum dots \cite{cap_coup_1,cap_coup_2,cap_coup3,cap_coup_4,cap_coup_5}.   \\
\indent As discussed earlier, the system under investigation operates on the basis of Coulomb-coupling between $S_M$ and $S_G$.  Due to capacitive coupling between $S_G$ and $S_M$, any change in electron number in $S_{M(G)}$ changes the electrostatic energy of $S_{G(M)}$. Maximum conductance between the reservoirs $L$ and $R$ is thus achieved when the ground state of $S_G$ is unoccupied.  As demonstrated in Fig.~\ref{fig:Fig_2}(a), the ground states of $S_L,~S_M$, and $S_R$ are energetically aligned with each other when the ground state of the dot $S_G$ is unoccupied. This facilitates electronic tunneling between the dots and thus provides a path for current flow. Any tunneling into the ground state of $S_G$ (demonstrated in Fig.~\ref{fig:Fig_2}.b)  enhances the electrostatic energy in $S_M$, thereby destroying the ground state alignment of $S_L$ and  $S_R$ with the Coulomb-blocked ground state of $S_M$. This restricts electronic flow between the dots. Hence, maximum conductance is achieved when the ground state of the dot $S_G$ is unoccupied\color{black}. The average conductance between $L$ and $R$ is, thus,   dependent on the average occupancy probability of $S_G$.   Any change in temperature of the reservoir $G$ changes the average ground state occupancy probability of  $S_G$. This  affects the current flow between $L$ and $R$, resulting in a temperature-sensitive conductance. We will later demonstrate that the sign of sensitivity is dependent on the relative position of $\varepsilon_G$ compared to the Fermi energy.\\
\indent  To assess the enhancement in performance, system-I is compared with the simplest non-local thermometer. The simplest non-local thermometer consists  of two Coulomb-coupled quantum dots, and  is demonstrated in Fig.~\ref{fig:Fig_1}(b) as system-II. System-II  is originally based on the thermodynamic engine conceived earlier by S\'anchez \emph{et. al.} \cite{heatengine5}. It consists of two Coulomb-coupled quantum dots $S_1$ and $G_1$ \cite{sensor_qdot}. $S_1$ is electrically coupled to the reservoirs $L$ and $R$ respectively, while $G_1$ is electrically coupled to the reservoir $G$. The dots $S_1$ and $G_1$ are capacitively coupled with mutual charging energy $U_m$.  The thermoelectric open circuit voltage based thermometry performance of system-II was analyzed  in detail by Zhang, \emph{et. al.} \cite{sensor_qdot}\color{black}.\\ 
\indent As stated earlier, we assume constant system-to-reservoir coupling between the reservoirs and quantum dots. Although Coulomb-coupled thermoelectric systems have been proposed as thermometers, such set-ups demand asymmetric step-like system-to-reservoir coupling for optimal operation \cite{thermalrefrigerator5,sensor_qdot}. Realistic implementation of such interfaces with sharp step-like coupling poses a huge fabrication challenge and cannot be achieved in a practical scenario. The set-up, considered in this paper, circumvents the need for any change in the system-to-reservoir coupling for optimal operation and hence paves the way towards the realistic fabrication of Coulomb-coupled system based thermometer. Besides, the symmetric design of the quadruple quantum dot thermometer with respect to the reservoirs $L$ and $R$, in addition to an energy-independent system-to-reservoir coupling, nullifies any fluctuation in system-current due to non-local thermoelectric action \cite{aniket_nonlocalheat}. This makes the system sensitivity independent of the average current path temperature (discussed later). \\ 
\indent We employ density matrix formulation to study the transport phenomena and performance of the set-up under consideration  (derived in Appendix \ref{app_a}). In Fig.~\ref{fig:Fig_1}(a), we denote the reservoir-to-system coupling by  $\gamma_c$ while, $t_L$ and $t_R$ represents the inter-dot tunnel coupling amplitudes. The fluctuation in  electrostatic energy of the entire system consisting of four quantum dots, due to electronic tunneling into the ground states,  {can be written as \cite{aniket_nonlocalheat, sispad}}:
 \begin{widetext}
	\begin{equation}
	U(n_{S_{L}},n_{S_{M}},n_{S_{R}},n_{S_G})=\sum_{x }U^{self}_{x}\left(n_{x}^{tot}-n_x^{eq}\right)^2   +\sum_{(x_{1},x_{2})}^{x_1 \neq x_2} U^m_{x_1,x_2}\left(n_{x_1}^{tot}-n_{x_1}^{eq}\right)\left(n_{x_2}^{tot}-n_{x_2}^{eq}\right), 
	\end{equation}
\end{widetext}
$n_x^{tot}$  being the total  electron number, and $U^{self}_x=\frac{q^2}{C^{self}_{x}}$ is the electrostatic potential energy due to self-capacitance $C^{self}_{x}$ (with the surrounding leads) of   the  dot `$x$' (details given in  Appendix \ref{app_a}). $U^m_{x_1,x_2}$ is the electrostatic energy due to  Coulomb-coupling between two neighbouring quantum dots, and $n_x^{eq}$ is the overall equilibrium number of electrons present in dot $x$ at $0$K and is determined by the minimum attainable electrostatic energy of the system. $n_x=n_x^{tot}-n_x^{eq}$, hence, defines the total number of electrons added in the ground state of the dot $x$ because of stochastic thermal fluctuations from the reservoirs. Here, a minimal physics model is used to study the thermometer performance under the assumption that the fluctuation in potential due self-capacitance of the quantum dots is much more compared to the average thermal voltage $kT/q$ or the applied bias voltage $V$, that is $U^{self}_x=\frac{q^2}{C^{self}_{x}}>> (kT,~qV)$. Thus, the electron occupancy probability or transport rate via the self-capacitance induced Coulomb-blocked energy state is negligibly small.  The analysis of the entire system of four quantum dots may hence be completed by limiting the maximum number of electrons in the ground state of each dot to one. Hence, the entire system investigation may be limited to sixteen multi-electron states, which we denote by the electron occupation number in the ground state of each quantum dot. Thus, a possible state of consideration in the system may be denoted  as $\ket{n_{S_L},n_{S_M},n_{S_R},n_{S_G}}=\ket{n_{S_L}}\tens{} \ket{n_{S_M}} \tens{} \ket{n_{S_R}} \tens{} \ket{n_{S_G}}$, where $n_{S_L},n_{S_M},n_{S_R},n_{S_G}\in (0,1)$,  denote the number of electrons present in the ground-states of $S_L,~S_M,~S_R$ and $S_G$ respectively. We also assume that the strength of  capacitive coupling between the dots, except for that between $S_M$ and $S_G$,  are negligible, such that, for all practical purposes under consideration, $U^m_{x,y} \approx 0$, for $(x,y) \neq (S_M,S_G)$. \\
\begin{figure}[!htb]
	\includegraphics[width=0.45\textwidth]{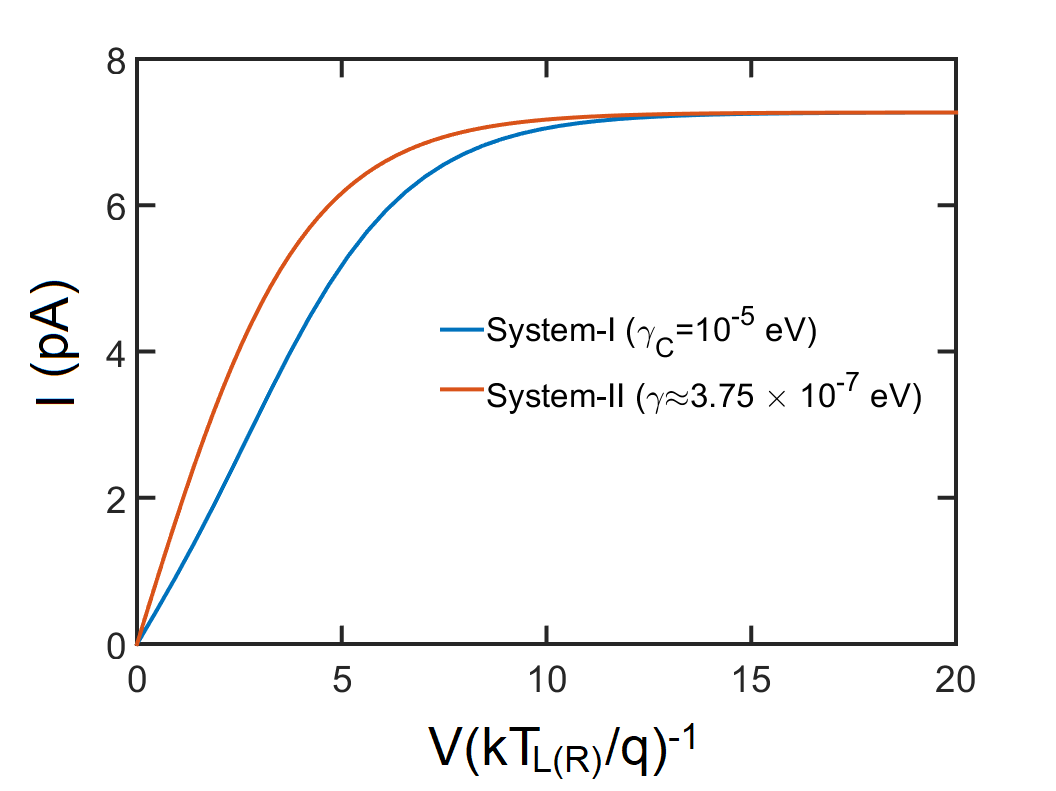}
	\caption{Benchmark of the maximum current of system-II with that of system-I. In case of system-I, $\gamma_c=10\upmu$eV. The maximum current of system-II becomes identical with that of system-I for $\gamma=0.375\upmu$eV (See Fig.~\ref{fig:Fig_1}). For calculating the maximum current through the systems, the ground states of the dots $S_G$ (for system-I) and $G_1$ (for system-II) are assumed to be empty (See Fig.~\ref{fig:Fig_1}) and the ground states of the dots $S_L,~S_M,~S_R,$ (for system-I) and $S_1$ (for system-II) are assumed to be aligned with the equilibrium Fermi potential. The temperature of the reservoirs are assumed to be $T_L=T_R=T_G=10$K.}
	\label{fig:Fig_3}
\end{figure}
 \begin{figure}[!htb]
	\includegraphics[width=0.5\textwidth]{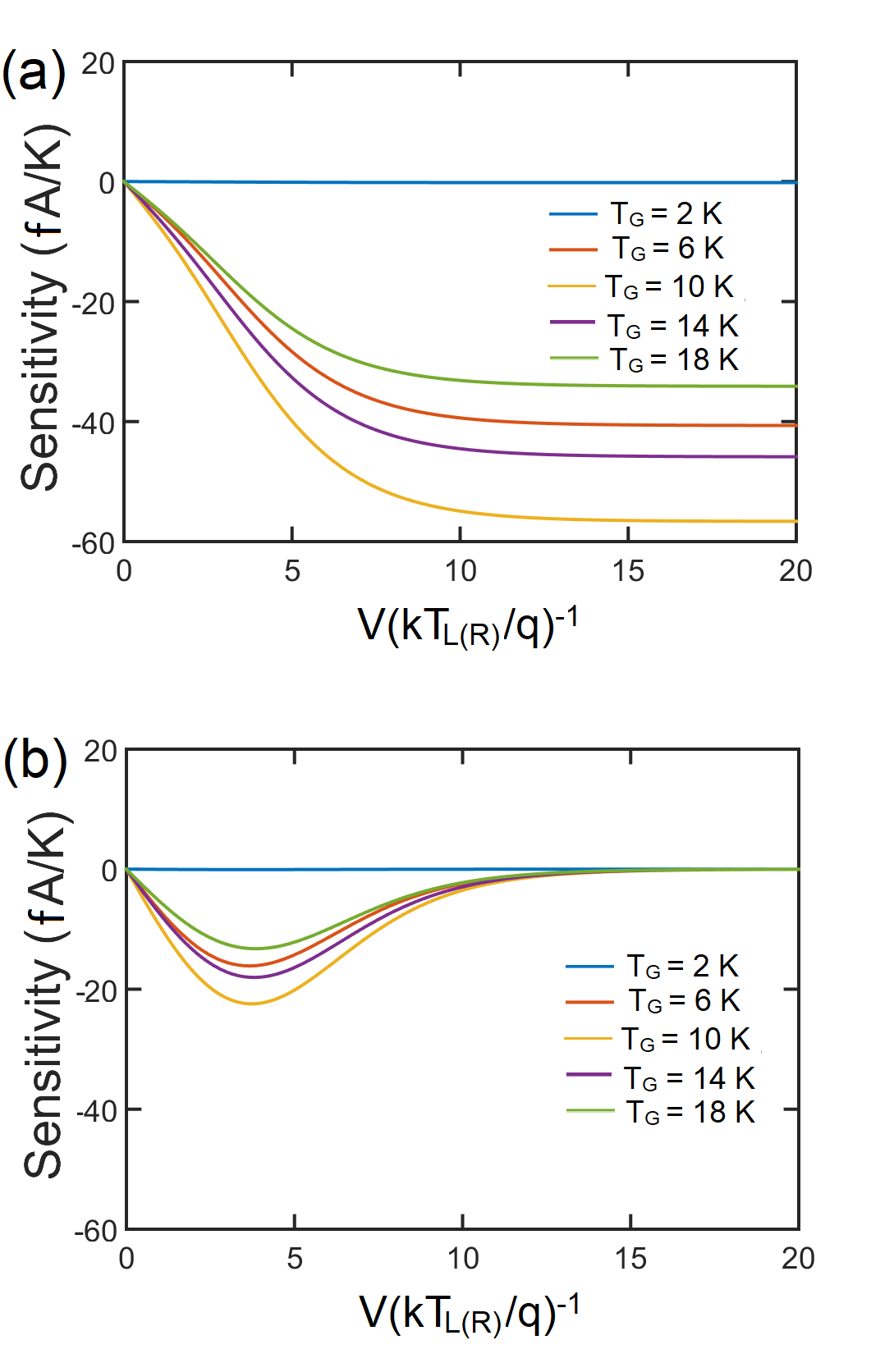}
	\caption{Variation of temperature sensitivity for (a) system-I and (b) system-II against voltage bias for various values of $T_{G}$ and fixed  $T_{L(R)}=10$K. The mutual   Coulomb-coupling energy is taken to be $U_m=2$meV$~(\approx 2.3209\frac{kT_{L(R)}}{q})$. We assume that the ground state energy levels are aligned to $\mu_0$, that is, $\varepsilon_L$=$\varepsilon_M$=$\varepsilon_R$=$\varepsilon_G$=$\mu_{0}$.}
	\label{fig:Fig_4}
\end{figure}
\begin{figure*}[!htb]
	\includegraphics[width=0.8\textwidth]{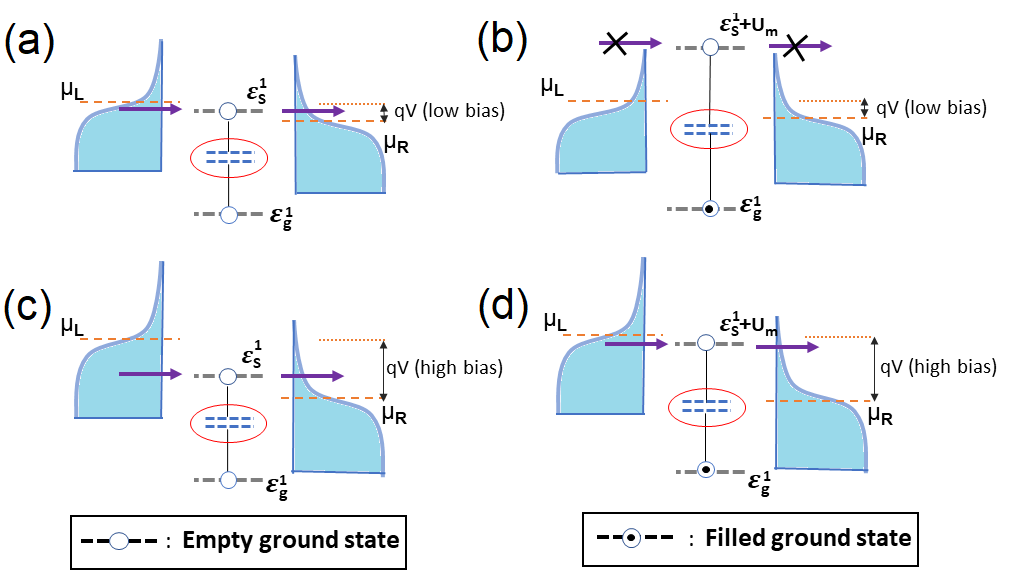}
	\caption{Electron transport through system-II in case of  low bias (top panel) and  high bias (bottom panel) situations.  The left and right panel respectively depict the transport through the ground state ($\varepsilon_{s}^1$) and the Coulomb-blocked state ($\varepsilon_{s}^1+U_m$) of $S_M$. In case of low bias (a) Electron can flow through the ground state $\varepsilon_s^1$. (b) However, no electrons can flow through the Coulomb-blocked level $\varepsilon_s^1+U_m$ due to absence of electrons in the reservoirs at that energy. (c, d) High bias populates the high energy states in the reservoir $L$ (connected to the negative pole of the power source). Thus electrons can now flow through both (c) the ground state $\varepsilon_s^1$ and, (d) the Coulomb-blocked state $\varepsilon_s^1+U_m$. Thus, current is maximized at sufficiently high bias resulting in \textit{zero} temperature sensitivity.}
	\label{fig:Fig_5}
\end{figure*} 
\begin{figure}[!htb]
	\includegraphics[width=0.5\textwidth]{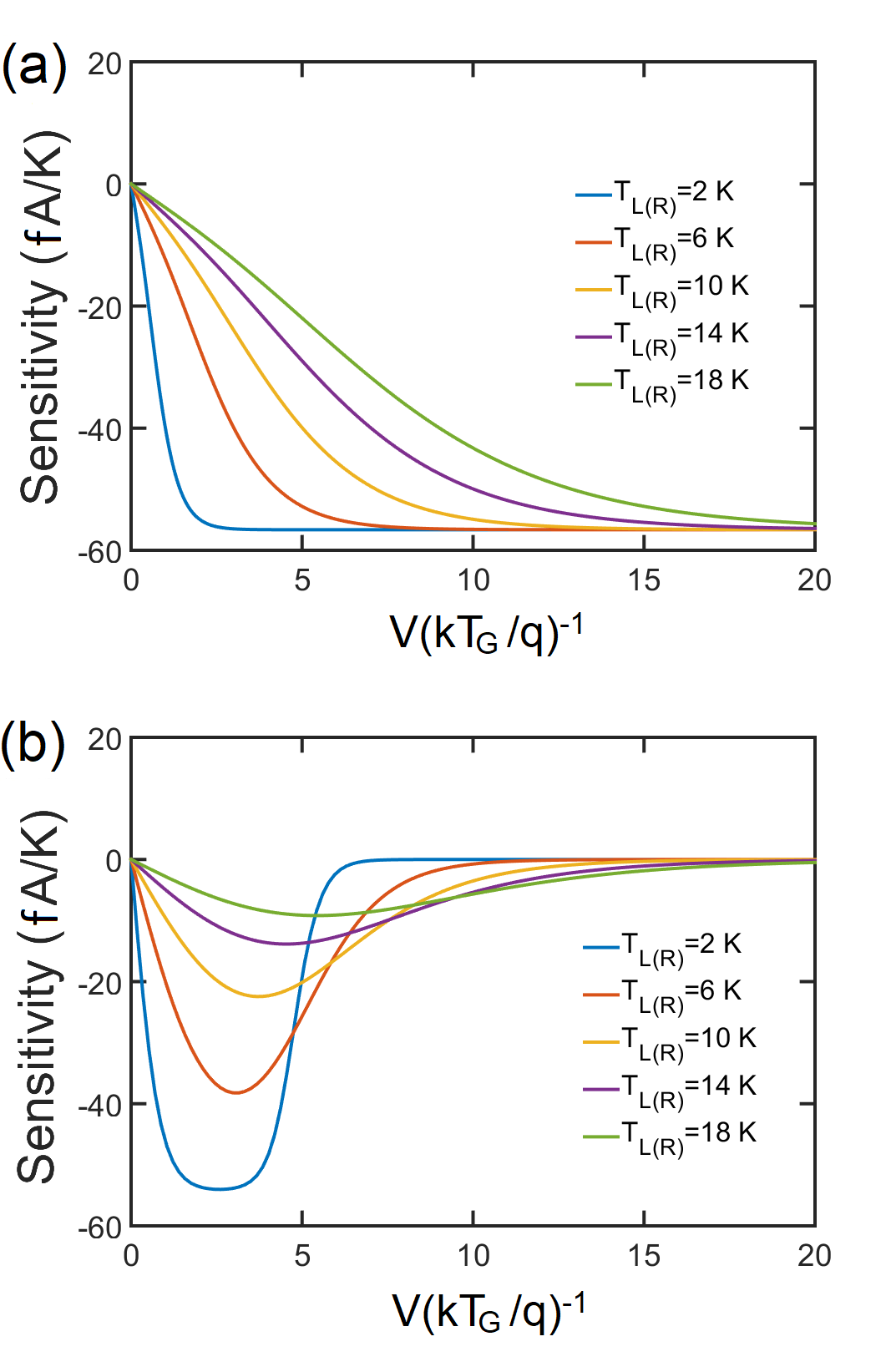}
	\caption{Variation of temperature sensitivity for (a) system-I and (b) system-II against voltage bias for various values of $T_{L(R)}$ and fixed remote reservoir $T_G=10$K and  Coulomb-coupling energy $U_m=2$meV$~(\approx 2.3209\frac{kT_G}{q})$. We assume that the ground state energy levels are aligned to $\mu_0$, that is, $\varepsilon_L$=$\varepsilon_M$=$\varepsilon_R$=$\varepsilon_G$=$\mu_{0}$.}
	\label{fig:Fig_6}
\end{figure} 
\begin{figure}[!htb]
	\includegraphics[width=0.5\textwidth]{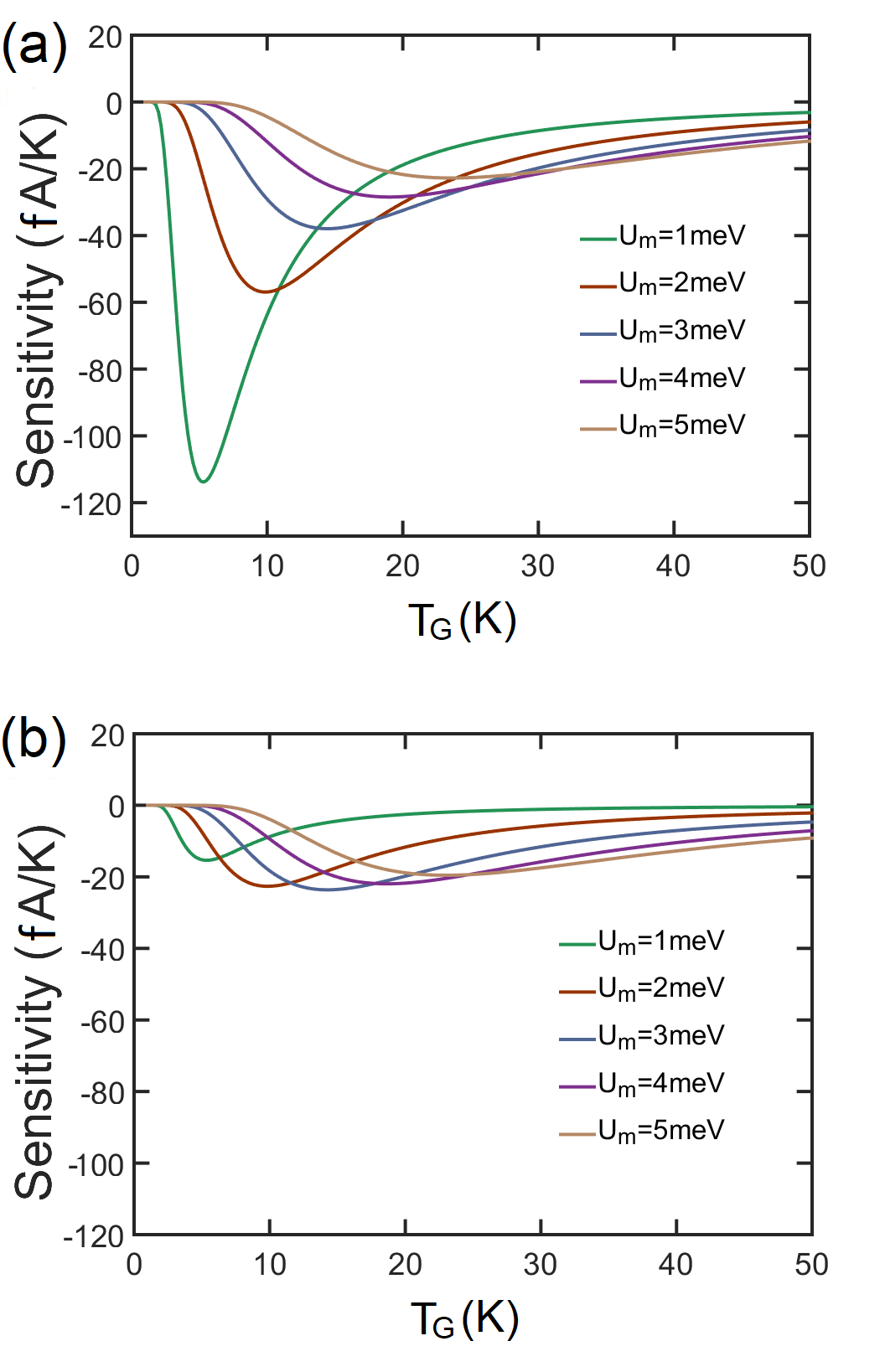}
	\caption{Sensitivity as a function of the remote reservoir temperature $T_G$ for various values of $U_m$ for  (a)  system-I and (b) system-II. The temperature of reservoirs $L$ and $R$ are fixed at $T_{L(R)}=10$K, while  the quantum dot ground states are aligned with the equilibrium Fermi energy, that is, $\varepsilon_L$=$\varepsilon_M$=$\varepsilon_R$=$\varepsilon_G$=$\mu_{0}$.}
	\label{fig:Fig_7}
\end{figure}
 \indent Due to capacitive coupling induced transport dependence among the ground states of $S_M$ and $S_G$, we treat this pair of dots as a  sub-system ($\varsigma_M$) of the entire system of four dots. The other two sub-systems $\varsigma_L$ and $\varsigma_R$ consist of the dots $S_L$ and  $S_R$, respectively.  The state probability of the sub-system  $\varsigma_M$ is denoted by $P_{i,j}^{\varsigma_M}$, where  $i$ and $j$ represent the electron number in the ground state of dot $S_M$ and $S_G$ respectively. $P_k^{\varsigma_{L(R)}}$, on the other hand, denotes the probability of occupancy of the dot $S_{L(R)}$. Under the condition that  $U_{S_M,S_G}$ is a few times greater than the ground state broadening due to system-to-reservoir coupling,  the inter-dot tunneling rates are maximized when either $\varepsilon_M$ or $\varepsilon_M+U_{S_M,S_G}$ coincides with the energy levels $\varepsilon_L$ and $\varepsilon_R$ (See Appendix \ref{app_a}).  For simplicity, we henceforth  represent $U_{S_M,S_G}$ as $U_m$. 	The state probabilities $P_{i,j}^{\varsigma_M}$ and  $P_k^{\varsigma_{L(R)}}$ are computed using density matrix formulation derived in Appendix \ref{app_a} (Eqs. \ref{eq:system_prob}), assuming quasi-equilibrium Fermi-Dirac statistics at the reservoirs. The probability of occupancy of the reservoirs at energy $\varepsilon$ is thus given by:
  \begin{equation}
 f_{\upsilon}(\varepsilon)=\left(1+exp\left\{\frac{\varepsilon-\mu_{\upsilon}}{kT_{\upsilon}}\right\}\right)^{-1},
 \end{equation} 
where $\mu_{\upsilon}$ and $T_{\upsilon}$ respectively represent the quasi-Fermi energy and temperature of the reservoir $\upsilon$. To analyze the  thermometer's performance, we use a voltage-controlled model, where a bias voltage $V$ is applied, with the positive and negative terminals being connected to $R$ and $L$, respectively. On the application of a bias voltage $V$, the different quasi-Fermi energy of the reservoirs under quasi-equilibrium condition may be written as $\mu_G=\mu_0, ~\mu_L=\mu_0+qV/2$ and $\mu_R=\mu_0-qV/2$, where $\mu_G, ~\mu_L$ and $\mu_R$ denote the quasi-Fermi energy of the reservoirs $G,~L$ and $R$ respectively and $\mu_0$ is the equilibrium Fermi energy of the entire system under consideration. Upon calculating the sub-system state probabilities, the electronic  current flowing into (out-of) the left (right) reservoirs can be written as (Appendix \ref{app_a}):
\small
\begin{align}
I_{L(R)}=\frac{q^2}{h}\gamma_c \left\{P^{\varsigma_{L(R)}}_0 f_{L(R)}(\varepsilon_{L(R)})-P^{\varsigma_{L(R)}}_1 \{1-f_{L(R)}(\varepsilon_{L(R)})\}\right\},
\label{eq:current1}
\end{align}
\normalsize
 where $\gamma_c$ denotes the reservoir-to-system coupling in $eV$ (Fig.~\ref{fig:Fig_1}.a). Although cryogenic thermometers, with sensitivity based on thermoelectric open-circuit voltage measurement, have been proposed and explored in literature \cite{thermalrefrigerator5,sensor_qdot}, the sensitivity  in such thermometers is dependent on both the average current path temperature, as well as, the temperature of the remote target reservoir. Such dependence of the sensitivity on the average temperature of the current path is an undesirable effect and may lead to an erroneous prediction of the target reservoir temperature ($T_G$)\color{black}.  In this aspect, the energy-independent system-to-reservoir coupling, in conjugation with the symmetrical design of the quadruple quantum dot thermometer with respect to the reservoirs $L$ and $R$ nullifies any non-local thermoelectric open-circuited voltage. With an applied external voltage bias, we hence define the thermometer sensitivity as the rate of change of electronic current between $L$ and $R$ with the remote reservoir temperature $T_G$.  We thus define the thermometer sensitivity as:
\begin{equation}
	Sensitivity ~(\chi) =  \left(\frac{dI}{dT_G}\right),
\end{equation}
 where $I=I_L=-I_R$, since outflow of electrons from reservoir $L$ implies inflow of electrons into reservoir $R$ (and vice-versa). In addition, to enhance the magnitude of temperature sensitivity, it is also important to impart sensitivity robustness against any fluctuation in voltage and current path temperature, which may vary by a large fraction due to Joule heating (particularly in the low-temperature regime)\color{black}. In this aspect, it should be noted that the thermoelectric insensitivity of system-I, in addition to energy filtering via the dots $S_L$ and $S_R$, renders a current-path temperature-independent thermometer sensitivity (discussed later). \\
\section{Results}\label{results}
\begin{figure}[!htb]
	\includegraphics[width=.506\textwidth]{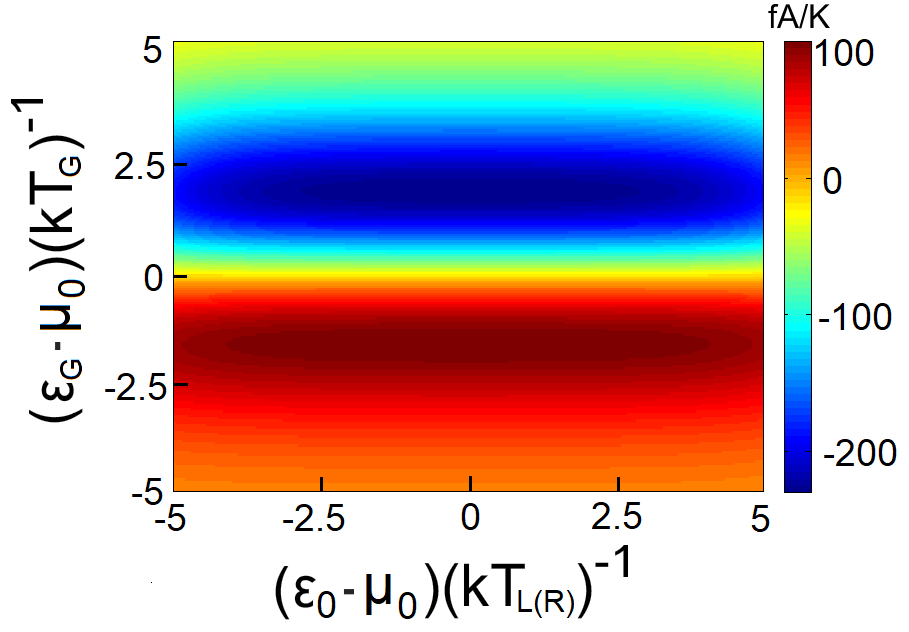}
	\caption{Colour plot depicting the performance of the system-I with variation in the ground states  for $U_m=2$meV$~(\approx 2.3209\frac{kT}{q})$ and  $V=0.0129$V$~(\approx 15\frac{kT}{q})$. The ground state of the dots $S_L$, $S_M$ and $S_R$ are aligned with each other as $\varepsilon_{L}=\varepsilon_{M}=\varepsilon_{R}=\varepsilon_{0}$. The temperature of the reservoirs $L,~R$ and $G$  are chosen as $T_L=T_R=T_G=10$K.}
	\label{fig:Fig_8}
\end{figure}

\indent In this section, we investigate the performance and the regime of operation of the quadruple quantum dot thermometer (system-I). In addition, we also conduct a performance comparison between system-I and system-II in terms of temperature sensitivity. To investigate the quadruple quantum dot thermometer (system-I), we choose the system-to-reservoir coupling to be $\gamma_c=10\upmu$eV. Such order of coupling parameter also correspond to realistic experimental values in Ref. \cite{heatengine13}, where the system-to-reservoir coupling was calculated, from experimental data, to lie in the range of $20\sim 50\upmu$eV. \color{black}   In addition, to validate the equations derived via density matrix formulation (Appendix \ref{app_a}) for the set-up under investigation, we choose $t_L=t_R=0.1\gamma_c$. Such values of the coupling parameters also signify the weak coupling limit and restrict the electronic transport in the sequential tunneling regime where the effects of co-tunneling and higher-order tunneling processes can be neglected. Unless stated, the temperature of the reservoirs $L$ and $R$ are assumed to be $T_L=T_R=10$K. In addition, to compare the performance of system-I to that of system-II, we choose $\gamma \approx 0.375\upmu$eV (Fig.~\ref{fig:Fig_1}.b), which in-turn causes identical maximum current for both the systems (See Fig.~\ref{fig:Fig_3}). This results in a comparison of two systems with identical maximum power dissipation (for a given voltage bias greater than a few $k_BT_{L(R)}$).\\
\indent \textbf{Performance comparison between system-I and system-II:} Fig.~\ref{fig:Fig_4}(a) demonstrates the variation of system-I sensitivity with the applied bias, for $U_m=2$meV, $\varepsilon_L=\varepsilon_M=\varepsilon_R=\varepsilon_G=\mu_0$  and different values of remote reservoir temperature $T_G$. We note that the sensitivity magnitude increases with bias $V$, up to the point of saturation. Such an initial increase in temperature sensitivity with bias voltage is due to an increase in current with $V$ (Fig. \ref{fig:Fig_3}).  Besides, the energy filtering by the dots $S_L$ and $S_R$ prohibits any electronic flow via the Coulomb-blockaded level $\varepsilon_M+U_m$ causing a constant temperature sensitivity as the bias voltage exceeds a few $kT_{L(R)}$. An increase and subsequent saturation in the total current marks identical behaviour for the system sensitivity with applied bias. We note that for $U_m=2$meV, the system  sensitivity is zero for $T_G=2$K over the entire range of applied bias. This is because under the conditions $T_G=2$K, $\varepsilon_L=\varepsilon_M=\varepsilon_R=\varepsilon_G=\mu_0$, and $U_m=2$meV, the temperature induced change in occupancy probability of both the ground state $\varepsilon_{G}$ and the Coulomb-blocked state $\varepsilon_{G}+U_m$ are zero, which results in \textit{zero} temperature sensitivity. With an increase in $T_G$, the temperature induced change in occupancy probability acquires a finite value resulting in \textit{non-zero} thermometer sensitivity (Appendix \ref{app_b}). {Although not demonstrated here, a finite system sensitivity for $T_G\leq 2$K can be achieved for lower values of $U_m$ } or by pinning the ground state of $S_G$ just above $\mu_0$. Fig. \ref{fig:Fig_4}(b) demonstrates the variation in sensitivity with applied bias for system-II. We note that system-I demonstrates a higher maximum sensitivity compared to system II. This is due to the fact that in system-I, the dots $S_L$ and $S_R$ act as energy filters, thereby prohibiting appreciable current through the Coulomb-blocked level $\varepsilon_{M}+U_m$. This results in a drastic current reduction when the ground state $S_G$ is occupied.   We also note from Fig.~\ref{fig:Fig_4}(b) that the sensitivity of system-II varies over a range of applied bias and reduces to zero when the applied bias is increased. Such a behaviour implies that the system is prone to voltage noise.  The decrease in sensitivity in system-II with an increase in bias voltage beyond a certain limit occurs because thermometry in Coulomb-coupled systems arises from the blockage of electronic flow when the ground state of $G_1$ is occupied. When the applied bias is gradually increased, the current through the system-II increases, which increases the sensitivity magnitude. The sensitivity increases till current through the Coulomb-blocked level $\varepsilon_s^1$ is negligible, such that an increase in occupancy probability of $G_1$ decreases the current through $S_1$.  On further increase in applied bias voltage,  electrons can tunnel into (out of) the Coulomb-blocked energy level as $\mu_L$  approaches  $\varepsilon_s^1+U_m$, which gradually reduces the temperature sensitivity. When the applied bias is sufficiently high, such that $f_L(\varepsilon_s^1+U_m)\approx 1$ and $f_R(\varepsilon_s^1)\approx 0$,  current through the system is maximized, prohibiting any further change in current due to variation in remote reservoir temperature $T_G$. This results in zero temperature sensitivity for system-II at high values of applied bias. This phenomena is schematically demonstrated in Fig.~\ref{fig:Fig_5}, where Fig.~\ref{fig:Fig_5}(a) and (b) depict that electrons can flow through the ground state $\varepsilon_s^1$, but not via the Coulomb-blocked level $\varepsilon_s^1+U_m$. This is due to the absence of electrons in the reservoirs at energy $\varepsilon_s^1+U_m$. Under high bias, however, the high energy states in the reservoir $L$ are populated giving rise to electron flow via both the ground state $\varepsilon_s^1$, and the Coulomb-blocked level $\varepsilon_s^1+U_m$. This is demonstrated in Fig.~\ref{fig:Fig_5}(c) and (d). Thus under sufficient high-bias, the current is maximized, resulting in \emph{zero} temperature sensitivity. \color{black} ~The sensitivity of  system-I, as demonstrated in Fig.~\ref{fig:Fig_4}(a) on the other hand, remains constant when the bias voltage $V$ increases beyond $10\frac{kT_{L(R)}}{q}$, which makes system-I robust against noise at high values of applied bias. It should be noted that the constant high bias sensitivity in system-I arises from the energy filtering effect of $S_L$ and $S_R$, that prohibits electronic flow through the Coulomb-blocked level $\varepsilon_{M}+U_m$ (demonstrated in Fig.~\ref{fig:Fig_2}). Hence, even under high bias condition, an electron tunneling in the ground state of $S_G$ completely chokes the current, giving rise to a non-zero sensitivity. \color{black}\\
 \indent Fig.~\ref{fig:Fig_6} demonstrates the variation in temperature sensitivity with voltage bias for various values of $T_{L(R)}$. In particular, Fig.~\ref{fig:Fig_6}(a) demonstrates the temperature sensitivity for system-I.  We note that at low values of the applied voltage bias ($V$),  the system sensitivity is dependent on $T_{L(R)}$. This is due to the fact that at low values of applied bias, the magnitudes of $f_L(\varepsilon_L)$ and $f_R(\varepsilon_R)$ are dependent on $T_L$ and $T_R$ respectively, causing the sensitivity to vary with $T_{L(R)}$. However, when the bias voltage is sufficiently increased, the magnitudes of $f_L(\varepsilon_L)$ and $f_R(\varepsilon_R)$ saturate to $f_L(\varepsilon_L)=1$ and $f_R(\varepsilon_R)=0$, resulting in   sensitivity saturation to a unique value for different  of $T_{L(R)}$ saturates\color{black}.  Thus for sufficiently high bias $V$,  system-I is robust against any variation of the average  current path temperature. On the other hand, for system-II, as demonstrated in Fig.~\ref{fig:Fig_6}(b), there is no value of the applied bias $V$ for which the system sensitivity is independent of $T_{L(R)}$. Thus in the case of system-II, the sensitivity is dependent on the average temperature of the current path. This implies that system-II is not robust against variation in the average current path temperature .  For the same reason as Fig.~\ref{fig:Fig_4}(b), the system-II sensitivity, demonstrated in Fig.~\ref{fig:Fig_6}(b), deteriorates to \emph{zero} at high values of the applied bias. It should again be noted that similar to Fig.~\ref{fig:Fig_4}(a), the robustness of system-I against variation in the temperature of current path, at high values of applied bias $V$, should be attributed to the energy filtering effect of $S_L$ and $S_R$. As stated earlier, the energy filtering effect of the dots $S_L$ and $S_R$ prevent any electronic flow via the Coulomb-blocked level $\varepsilon_{M}+U_m$ (demonstrated in Fig.~\ref{fig:Fig_2}). This preserves the provision of conductance change due to electronic tunneling in $S_G$.\\ \color{black}
\indent  Fig.~\ref{fig:Fig_7} demonstrates the variation of saturation sensitivity (at high bias) with remote reservoir temperature $T_G$ for different values of the Coulomb-coupling energy ($U_m$). In particular, Fig.~\ref{fig:Fig_7} (a) demonstrates the variation of system-I saturation sensitivity with $T_G$. To calculate the maximum or saturation sensitivity, we choose  $V=0.0129$V or $\approx 15\frac{kT_{L(R)}}{q}$. We note that at a lower target reservoir temperature $T_G$, the system sensitivity is higher for lower values of $U_m$. Clearly, for a particular $U_m$, there exists an optimal remote reservoir temperature $T_G=T_G^{opt}$ at which the system sensitivity is maximum.  We also note that the maximum sensitivity at a higher temperature demands a  higher value of $U_m$. This optimal $T_G$ ($T_G^{opt}$) as a function of $U_m$ can be given as (Appendix \ref{app_b}):
\begin{equation}
	T_G^{opt}=(2.399)^{-1}\frac{U_m}{ k}\label{eq:1}
\end{equation}
From Eqn. \ref{eq:1}, it can be noted that the value of $T_G^{opt}$ increases with increase in $U_m$, a trend which is clearly clearly reflected in Fig.~\ref{fig:Fig_7}(a). Fig.~\ref{fig:Fig_7}(b) demonstrates the maximum sensitivity of system-II for different values of $U_m$  with a variation in the remote reservoir temperature. For computing the maximum sensitivity of system-II, we choose optimal bias voltage that results in maximum sensitivity (Appendix \ref{app_c}). It is clear that in the low temperature regime, system-I demonstrates a much  higher magnitude of sensitivity compared to system-II. Thus, system-I demonstrates an overall superior performance compared to system-II in terms of high magnitude of temperature sensitivity and robustness against fluctuations in voltage bias and average temperature of the current path. \\
\indent \textbf{Regime of operation:} Fig.~\ref{fig:Fig_8} demonstrates the operational regime  of system-I in terms of the ground state configurations of the quantum dots for  $U_m=2$meV$~(\approx 2.3209\frac{kT}{q})$, $V=0.0129$V$~(\approx  15\frac{kT}{q})$ and $T_L$=$T_R$=$T_G=10$K. While investigating the regime of operation, the ground states of $S_L,~S_M$ and $S_R$ are assumed to be aligned as $\varepsilon_L=\varepsilon_M=\varepsilon_R=\varepsilon_0$. We note that the sensitivity magnitude is maximum when $\varepsilon_{G}$ lies a few $kT_G$ above the equilibrium Fermi energy. This can be explained as follows. When $\varepsilon_{G}-\mu_0=0$, a change in $T_G$ results in a variation of stochastic fluctuation at the Coulomb-blocked level $\varepsilon_{G}+U_m$. There is, however, no variation in stochastic fluctuation of the ground state $\varepsilon_{G}$. As the ground state $\varepsilon_{G}$
rises above $\mu_0$, a change in $T_G$ affects the stochastic fluctuation of both the ground state $\varepsilon_{G}$ and the Coulomb-blocked level $\varepsilon_{G}+U_m$, which increases the system sensitivity to $T_G$. When $\varepsilon_{G}$ is further increased above $\mu_0$, both the function $\frac{df_G(\varepsilon_{G})}{dT_G}$ and $\frac{d}{dT_G}f_G(\varepsilon_{G}+U_m)$ reduces causing a drop in the overall sensitivity. We thus note that the regime of optimal sensitivity, in terms of $\varepsilon_{G}$ lies in the range of  $1\sim 3 kT_G$ above $\mu_0$. We also note that the sensitivity, as $\varepsilon_{G}$ goes below $\mu_0$, becomes positive. This is due to the fact that as $\varepsilon_{G}$ gradually goes below $\mu_0$, an increase in $T_G$ causes a decrease in the ground state occupancy of $S_G$ ($\frac{df_G(\zeta)}{dT_G}<0$ when $\zeta<\mu_0$), resulting in an increase in current flow with an increase in temperature. Fig.~\ref{fig:Fig_8} also demonstrates that system-I offers good temperature sensitivity over a wide range of $\varepsilon_{0}$, which is $-4kT_{L(R)}\lessapprox \varepsilon_{0} \lessapprox  4kT_{L(R)}$.\\
\begin{figure}[!htb]
	\centering
	\includegraphics[width=.5\textwidth]{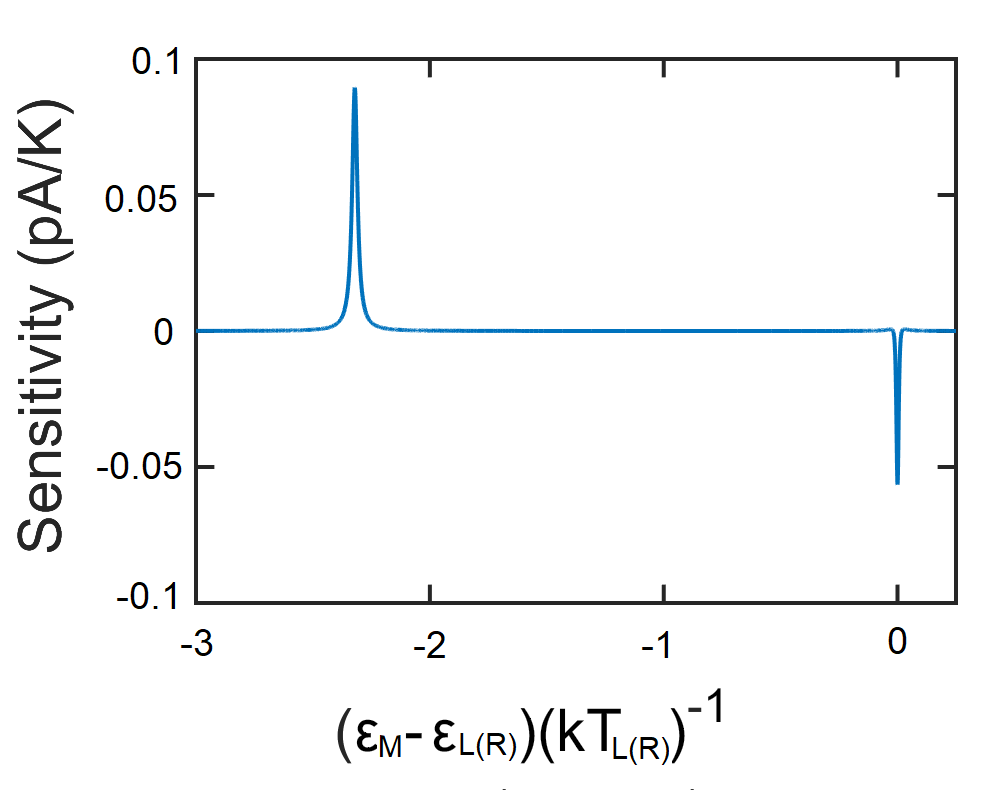}
	\caption{Plot of sensitivity of system-I with variation in the energy level $\varepsilon_M$. The Coulomb-coupling energy is assumed to be $U_m=2$meV$~(\approx 2.3209\frac{kT}{q})$. The bias V is chosen as $V=0.0129$V$~(\approx 15\frac{kT}{q})$. Temperature of the reservoirs $L$, $R$ and $G$ are taken as $T_L$=$T_R$=$T_G=10$K. The other ground state energy levels are aligned to $\mu_0$, that is, $\varepsilon_L$=$\varepsilon_R$=$\varepsilon_G$=$\mu_{0}$}
	\label{fig:Fig_9}
\end{figure}
\indent The variation in sensitivity with the ground state configuration of  $S_{M}$ is demonstrated in  Fig.~\ref{fig:Fig_9}. In this case the ground states of the dots $S_L$, $S_R$ and $S_G$ are kept fixed at  $\varepsilon_L$=$\varepsilon_R$=$\varepsilon_G$=$\mu_{0}$, while  $\varepsilon_{M}$ is varied to obtain the positions of optimal sensitivity. As already noted in the previous discussions, the sensitivity is negative and high when $\varepsilon_M$ coincides with $\varepsilon_{L(R)}$. This is expected since the current through the system is maximum when the ground states of the dots $S_L$, $S_M$ and $S_R$ are aligned with each other. The negative value of sensitivity arises from the fact that an increase temperature results in an increase in electron occupation probability in the ground state of the dot $S_G$ and thus decreases the current through the system. A misalignment in the ground state of $S_M$ with respect to the ground states of $S_{L(R)}$ results in a sharp deterioration in system sensitivity. Interestingly, we also note a positive peak in sensitivity around the $\varepsilon_M-\varepsilon_{L(R)}\approx 2.3\frac{kT}{q}$, which is equal to the value of $U_m$ in this case.  When $\varepsilon_{M}+U_m=\varepsilon_{L(R)}$, the  ground states $\varepsilon_{L(R)}$ is aligned with the Coulomb-blocked level $\varepsilon_{M}+U_m$. Hence, an electron tunneling into the ground state of $S_G$ increases the current flow through the system.  When $\varepsilon_G-\mu_0=0$, an increase in temperature increases the occupancy probability of the Coulomb-blockaded level $\varepsilon_G+U_m$, which in turn enhances the current flow. This in turn results in a positive sensitivity peak at $\varepsilon_M+U_m=\varepsilon_{L(R)}$. We thus note two possible configurations of the ground state of $S_M$, which results in finite optimal temperature sensitivity of system-I. However, an exact alignment of the Coulomb-blockaded level $\varepsilon_M+U_m$ with the ground states $\varepsilon_{L(R)}$ may be difficult to achieve experimentally.  Hence,  the configuration $\varepsilon_{L}=\varepsilon_{M}=\varepsilon_{R}$ is more realistic and attractive from a practical perspective. \\
 \section{Conclusions}\label{conclusion}
\indent To conclude, in this paper, we have investigated a Coulomb-coupled system based quadruple quantum dot thermometer. The performance and regime of operation of the thermometer were then theoretically analyzed using the density matrix formulation. It was demonstrated that the thermometer under consideration demonstrates superior performance and robustness compared to a simple double dot-based thermometer. {In addition, the symmetrical design of the system with respect to the reservoirs results in decoupling of non-local thermoelectric effects due to a deviation in remote reservoir temperature from the average temperature of the current path \cite{aniket_nonlocalheat, sispad}}. In this paper, we have analyzed the thermometer in the weak coupling regime where the effects of co-tunneling and higher-order tunneling processes can be neglected, resulting in the sensitivity of the order of tens of fA/K. The sensitivity can be increased by a few orders of magnitude by tuning the various tunnel coupling amplitudes in the strong coupling regime. It would be interesting to investigate the effects of co-tunneling on the thermometer performance as the system is gradually tuned towards the strong coupling regime. In addition, an investigation on the effects of electron-phonon scattering on the thermometer performance also constitutes an interesting direction. Other possible design strategies for electrical thermometers employing Coulomb-coupled quantum dots is left for future investigation. Nevertheless, the quadruple quantum dot design investigated in this paper can be used to fabricate highly sensitive and robust non-local electrical thermometers.  

   \indent \textbf{Acknowledgments:} Aniket Singha would like to thank financial support from Sponsored Research and Industrial Consultancy (IIT Kharagpur) via grant no. IIT/SRIC/EC/MWT/2019-20/162, Ministry of Human Resource Development (MHRD),  Government of India  via Grant No.  STARS/APR2019/PS/566/FS under STARS scheme and Science and Engineering Research Board (SERB),  Government of India  via Grant No. 
   SRG/2020/000593 under SRG scheme.\\
  \indent \textbf{Data Availability:} The data that supports the findings of this study are available within the article
\appendix
\begin{widetext}
\section{Derivation of state probabilities and current equations via density matrix formulation } \label{app_a}
	The total electrostatic energy of the system shown in Fig.~\ref{fig:Fig_1} can be given by the equation:
	\begin{equation}
	U(n_{S_{L}},n_{S_{M}},n_{S_{R}},n_{S_G})=\sum_{x }U^{self}_{x}\left(n_{x}^{tot}-n_x^{eq}\right)^2   +\sum_{(x_{1},x_{2})}^{x_1 \neq x_2} U^m_{x_1,x_2}\left(n_{x_1}^{tot}-n_{x_1}^{eq}\right)\left(n_{x_2}^{tot}-n_{x_2}^{eq}\right),
	\end{equation}
where the symbols have already been defined in the main text. As stated earlier, a possible state of the system may be denoted by $\ket{n_L,n_M,n_R,n_G}$, where $n_L,n_M,n_R,n_G \in (0,1)$. With a slight abuse of notations, to make the equations compact, I rename the states as $\ket{0,0,0,0}\rightarrow \ket{1}$, $\ket{0,0,0,1}\rightarrow \ket{2}$, $\ket{0,0,1,0}\rightarrow \ket{3}$, $\ket{0,0,1,1}\rightarrow \ket{4}$, $\ket{0,1,0,0}\rightarrow \ket{5}$, $\ket{0,1,0,1}\rightarrow \ket{6}$, $\ket{0,1,1,0}\rightarrow \ket{7}$, $\ket{0,1,1,1}\rightarrow \ket{8}$, $\ket{1,0,0,0}\rightarrow \ket{9}$, $\ket{1,0,0,1}\rightarrow \ket{10}$, $\ket{1,0,1,0}\rightarrow \ket{11}$, $\ket{1,0,1,1}\rightarrow \ket{12}$, $\ket{1,1,0,0}\rightarrow \ket{13}$, $\ket{1,1,0,1}\rightarrow \ket{14}$, $\ket{1,1,1,0}\rightarrow \ket{15}$, $\ket{1,1,1,1}\rightarrow \ket{16}$. Under this representation and the assumptions already stated in the main text, the Hamiltonian of the entire system (excluding the reservoirs) can be written as:
	\begin{multline}
H=\sum_{j=1}^{16}\epsilon_j \ket{j}\bra{j}+t_L\left\{\ket{5}\bra{9}+\ket{6}\bra{10}+\ket{7}\bra{11}+\ket{8}\bra{12} +~h.c. \right\}+t_R\left\{\ket{3}\bra{5}+\ket{4}\bra{6}+\ket{11}\bra{13}+\ket{12}\bra{14}+~h.c.\right\},  \\ 
\label{eq:hamiltonian}
\end{multline}
where $h.c.$ denotes  hermitian conjugate. In the above equations,  $\epsilon_x$ is the total energy of the state system in state $\ket{x}$, assuming the state $\ket{0,0,0,0}$ or $\ket{1}$ to be the vacuum state. Hence, the different values of $\epsilon_x$ may be expressed as:
$\epsilon_1=0;$ 
$\epsilon_2=\varepsilon_G;$
$\epsilon_3=\varepsilon_R;$
$\epsilon_4=\varepsilon_R+\varepsilon_G;$
$\epsilon_5=\varepsilon_M;$
$\epsilon_6=\varepsilon_G+\varepsilon_M+U_m;$
$\epsilon_7=\varepsilon_M+\varepsilon_R;$
$\epsilon_8=\varepsilon_G+\varepsilon_M+\varepsilon_R+U_m;$
$\epsilon_9=\varepsilon_L;$
$\epsilon_{10}=\varepsilon_L+\varepsilon_G;$
$\epsilon_{11}=\varepsilon_L+\varepsilon_R;$
$\epsilon_{12}=\varepsilon_L+\varepsilon_R+\varepsilon_G;$
$\epsilon_{13}=\varepsilon_L+\varepsilon_M;$
$\epsilon_{14}=\varepsilon_L+\varepsilon_G+\varepsilon_M+U_m;$
$\epsilon_{15}=\varepsilon_L+\varepsilon_M+\varepsilon_R;$
$\epsilon_{16}=\varepsilon_L+\varepsilon_G+\varepsilon_M+\varepsilon_R+U_m$. Under the assumption that the inter-dot coupling is much weaker than the system to reservoir coupling, the temporal evolution of the density matrix elements, in the regime of dominant sequential transport between the reservoir and the system can be given by a set of modified Liouville equations \cite{master_eq_1,master_eq_2,master_eq_3,sispad,master_eq_4,master_eq_5,master_eq_6}:
\begin{eqnarray}
\frac{\partial \rho_{\eta \eta}}{\partial t}=-i[H,\rho]_{\eta \eta}-\sum_{\nu} \Gamma_{\eta \nu}\rho_{\eta \eta}+\sum_{\delta}\Gamma_{\delta \eta }\rho_{\delta \delta} \nonumber \\
\frac{\partial \rho_{\eta \beta}}{\partial t}=-i[H,\rho]_{\eta \beta}-\frac{1}{2}\sum_{\nu} \Big(\Gamma_{\eta \nu}+\Gamma_{\beta \nu }\Big) \rho_{\eta \beta},
\label{eq:time_derivate}
\end{eqnarray}
where $\rho_{\eta,\eta}=\bra{\eta}\rho \ket{\eta}$ and $\rho_{\eta,\beta}=\bra{\eta}\rho \ket{\beta}$, $\rho$ being the density matrix operator. In the above equation, $[A,B]$ denotes the commutator of the operators $A$ and $B$, while $\Gamma_{\eta \nu}$ denotes the rate of transition from state $\nu$ to state $\eta$ due to electronic transitions via system-to-reservoir coupling. $\Gamma_{\eta \nu}$ can be given by
\begin{equation}
\Gamma_{\eta \nu}=\gamma_c \times f_{\upsilon}(\epsilon_{\eta}-\epsilon_{\nu}-\mu_{\upsilon}),
\end{equation} 
where the electronic transition is driven by the reservoir $\upsilon$.
We now write the equation for the density matrix element $\rho_{5,5}$ from Eq. \eqref{eq:time_derivate}, assuming steady state.
	\begin{equation}
0=\dot{\rho}_{5,5}=i t_L\left\{\rho_{5,9}-\rho_{9,5}\right\}+i t_R\left\{\rho_{5,3}-\rho_{3,5}\right\}+ \sum_{j=6,7,13}\Big\{\Gamma_{j,5}\rho_{j,j}-\rho_{5,5}\Gamma_{5,j}\Big\}
\label{eq:55}
\end{equation}
A change in the state $\ket{5}$ of the system can occur due to external system-to-reservoir coupling or due to inter-dot tunnelling. Inter-dot tunnelling,  can drive the system from state $\ket{5}$ to $\ket{3}$ or $\ket{9}$. In Eq.~\ref{eq:55}, the first term accounts to inter-dot tunnelling between the dots $S_L$ and $S_M$, the second term accounts for the inter-dot tunnelling between dot $S_M$ and $S_R$, while the third term accounts for the change in system state due to system-to-reservoir coupling. The expression of the non-diagonal density matrix elements $\rho_{3,5}=\rho^*_{5,3}$, under steady state,  can be derived  from Eq. \eqref{eq:time_derivate} as:
	\begin{equation}
0=\dot{\rho}_{3,5}=\dot{\rho^*}_{5,3}=i \rho_{3,5}\left\{\epsilon_5-\epsilon_3\right\}+i t_R\left\{\rho_{3,3}-\rho_{5,5}\right\}+it_L\rho_{3,9}-\frac{1}{2} \left\{\sum_{j=6,7,13}\Gamma_{5,j}+\sum_{j=1,4,11}\Gamma_{3,j}\right\}\rho_{3,5}
\end{equation}
\begin{equation}
{\rho}_{3,5}={\rho^*}_{5,3}=\frac{ t_R\left\{\rho_{3,3}-\rho_{5,5}\right\}+t_L\rho_{3,9}}{\left\{\epsilon_3-\epsilon_5\right\}-i\frac{1}{2} \Upsilon_{3,5}}
\label{eq:35}
\end{equation}
\begin{equation}
\Upsilon_{3,5}=\sum_{j=6,7,13}\Gamma_{5,j}+\sum_{j=1,4,11}\Gamma_{3,j}
\end{equation}
Similarly, 
\begin{equation}
{\rho}_{9,5}={\rho^*}_{5,9}=\frac{ t_L\left\{\rho_{9,9}-\rho_{5,5}\right\}+t_R\rho_{9,3}}{\left\{\epsilon_9-\epsilon_5\right\}-i\frac{1}{2} \Upsilon_{9,5}}
\label{eq:95}
\end{equation}
\begin{equation}
\Upsilon_{9,5}=\sum_{j=6,7,13}\Gamma_{5,j}+\sum_{j=1,10,11}\Gamma_{9,j}
\end{equation}
The numerator in Eqns.~\eqref{eq:35} and \eqref{eq:95} contains non-diagonal terms $\rho_{3,9}$ and $\rho_{9,3}$. Since, the Hamiltonian in Eq.~\eqref{eq:hamiltonian} doesn't contain a direct tunnelling matrix element between the states $\ket{9}$ and $\ket{3}$, it is imperative to understand the effect of this term on the electronic transport. Assuming steady state, from the non-diagonal elements of the density matrix, we get
\begin{eqnarray}
0=\dot{\rho}_{3,9}=\dot{\rho^*}_{9,3}=i \rho_{3,9}\left\{\epsilon_9-\epsilon_3\right\}+it_L\rho_{3,5}-it_R\rho_{5,9}-\frac{1}{2} \left\{\sum_{j=4,7,11}\Gamma_{3,j}+\sum_{j=1,10,11}\Gamma_{9,j}\right\}\rho_{3,9}
\end{eqnarray}
\begin{eqnarray}
{\rho}_{3,9}={\rho^*}_{9,3}=\frac{t_L\rho_{3,5}-t_R\rho_{5,9}}{\{\epsilon_3-\epsilon_9\}-\frac{i}{2}\Upsilon_{3,9}}
\label{eq:39}
\end{eqnarray}
\begin{equation}
\Upsilon_{3,9}=\sum_{j=4,7,11}\Gamma_{3,j}+\sum_{j=1,10,11}\Gamma_{9,j}
\end{equation}
The numerator in Eq.~\eqref{eq:39}, the element $t_L\rho_{3,5}$ accounts for co-tunnelling processes where simultaneous  transfer of electrons take place from dot $M$ to $R$  and from dot $L$ to $M$. Similarly, the element $t_R\rho_{5,9}$ accounts for another co-tunnelling process where the simultaneous transfer of  electrons take place  from dot $R$ to $M$ and from dot $M$ to $L$. For simplifying further derivations, we assume $t_L=t_R=t$. The values of $\rho_{3,5}$ and $\rho_{5,9}$ in Eq.~\eqref{eq:39} can be replaced from Eqns.~\eqref{eq:35} and \eqref{eq:95} to get an expansion containing the density matrix elements weighted by $t^2$. Repeated expansion of Eq.~\eqref{eq:39} by employing Eqn.~ \eqref{eq:35}, \eqref{eq:95} and  \eqref{eq:39} results in terms of higher order in $t$ and accounts for the higher order co-tunnelling processes. It can hence be noted that the terms in Eq. \eqref{eq:39} are  at-least of second order in $t$, when expanded in terms of the diagonal density-matrix elements. However, the expansion   $\rho_{3,5}$ and $\rho_{5,9}$ from Eqns.~\eqref{eq:35} and \eqref{eq:95} contains terms which are first order in $t$. Under the assumption that the system-to-reservoir coupling $\gamma_c$ is much higher than the  inter-dot tunnelling parameter $t$, the contribution of the term $\rho_{3,9}$ in electronic transport properties can be neglected with respect to the terms $\rho_{5,3}$ and $\rho_{5,9}$. This approximation accounts to neglecting the contributions of  co-tunnelling or higher order tunnelling  processes and  estimating the properties of the entire system via sequential transport or first order tunnelling  processes. Hence Eq.~\eqref{eq:35}  can be approximated as, 
\begin{equation}
{\rho}_{3,5}= {\rho^*}_{5,3}\approx\frac{ t_R\left\{\rho_{3,3}-\rho_{5,5}\right\}}{\left\{\epsilon_3-\epsilon_5\right\}-\frac{i}{2} \Upsilon_{3,5}}
\label{eq:350}
\end{equation}
\begin{equation}
{\rho}_{9,5}={\rho^*}_{5,9}\approx\frac{ t_L\left\{\rho_{9,9}-\rho_{5,5}\right\}}{\left\{\epsilon_9-\epsilon_5\right\}-\frac{i}{2} \Upsilon_{9,5}}
\label{eq:950}
\end{equation}
Putting the values of $\rho_{3,5}$ and $\rho_{9,5}$ from Eqns.~\eqref{eq:350} and \eqref{eq:950} in \eqref{eq:55}, we get,\\
\begin{equation}
\dot{\rho}_{5,5}=\tau_{3,5}\rho_{3,3}+\tau_{9,5}\rho_{9,9}-(\tau_{5,3}+\tau_{5,9})\rho_{5,5}+ \sum_{j=6,7,13}\Big\{\Gamma_{j,5}\rho_{j,j}-\rho_{5,5}\Gamma_{5,j}\Big\},
\label{eq:552}
\end{equation}
where $\tau_{3,5}=\tau_{5,3}=\tau_{3,5}=\frac{t^2\Upsilon_{5,3}}{(\epsilon_M-\epsilon_R)^2+\left(\frac{\Upsilon_{5,3}}{2}\right)^2}$ and $\tau_{5,9}=\tau_{9,5}=\frac{t^2\Upsilon_{5,9}}{(\epsilon_M-\epsilon_L)^2+\left(\frac{\Upsilon_{5,9}}{2}\right)^2}$. It is clear that in Eq.~\eqref{eq:552}, the term $\tau_{3(9),5}\rho_{3,3(9,9)}$ takes into account state transition from from $\ket{3(9)}$ to $\ket{5}$, due to  tunnelling from the the dot $S_{R (L)}$ to the dot $S_M$, while $\tau_{5,3(9)}\rho_{5,5}$ takes into account the transition from $\ket{5}$ to $\ket{3(9)}$ due to tunnelling from $S_M$ to $S_{L(R)}$. Thus, these terms denote the rate of inter-dot tunnelling.

\indent Similarly, the rate equations corresponding to the other diagonal elements of the density matrix, in the limit of negligible co-tunnelling phenomena can ultimately be derived as:
\begin{align}
&\dot{\rho}_{1,1}=\sum_{j=2,3,9}(\Gamma_{j,1}\rho_{j,j}-\Gamma_{1,j}\rho_{1,1}) \label{eq:set_first}\\
&\dot{\rho}_{2,2}=\sum_{j=1,4,10}(\Gamma_{j,2}\rho_{j,j}-\Gamma_{2,j}\rho_{2,2}) \\
&\dot{\rho}_{3,3}=\tau_{5,3}\rho_{5,5}-\tau_{3,5}\rho_{3,3} +\sum_{j=1,4,11}(\Gamma_{j,3}\rho_{j,j}-\Gamma_{3,j}\rho_{3,3}) \\
&\dot{\rho}_{4,4}=\tau_{6,4}\rho_{6,6}-\tau_{4,6}\rho_{4,4} +\sum_{j=2,3,12}(\Gamma_{j,4}\rho_{j,j}-\Gamma_{4,j}\rho_{4,4}) \\
&\dot{\rho}_{5,5}=\tau_{3,5}\rho_{3,3}+\tau_{9,5}\rho_{9,9}-(\tau_{5,3}+\tau_{5,9})\rho_{5,5}+ \sum_{j=6,7,13}\Big\{\Gamma_{j,5}\rho_{j,j}-\rho_{5,5}\Gamma_{5,j}\Big\}\\
&\dot{\rho}_{6,6}=\tau_{4,6}\rho_{4,4}+\tau_{10,6}\rho_{10,10}-(\tau_{6,4}+\tau_{6,10})\rho_{6,6} +\sum_{j=5,8,14}(\Gamma_{j,6}\rho_{j,j}-\Gamma_{6,j}\rho_{6,6}) \\
&\dot{\rho}_{7,7}=\tau_{11,7}\rho_{11,11}-\tau_{7,11}\rho_{7,7} +\sum_{j=5,8,15}(\Gamma_{j,7}\rho_{j,j}-\Gamma_{7,j}\rho_{7,7}) \\
&\dot{\rho}_{8,8}=\tau_{12,8}\rho_{12,12}-\tau_{8,12}\rho_{8,8} +\sum_{j=6,7,16}(\Gamma_{j,8}\rho_{j,j}-\Gamma_{8,j}\rho_{8,8}) \\
&\dot{\rho}_{9,9}=\tau_{5,9}\rho_{5,5}-\tau_{9,5}\rho_{9,9} +\sum_{j=1,10,11}(\Gamma_{j,9}\rho_{j,j}-\Gamma_{9,j}\rho_{9,9}) \\
&\dot{\rho}_{10,10}=\tau_{6,10}\rho_{6,6}-\tau_{10,6}\rho_{10,10} +\sum_{j=2,9,12}(\Gamma_{j,10}\rho_{j,j}-\Gamma_{10,j}\rho_{10,10})\\
&\dot{\rho}_{11,11}=\tau_{7,11}\rho_{7,7}+\tau_{13,11}\rho_{13,13}-(\tau_{11,7}+\tau_{11,13})\rho_{11,11} +\sum_{j=3,9,12}(\Gamma_{j,11}\rho_{j,j}-\Gamma_{11,j}\rho_{11,11}) \\
&\dot{\rho}_{12,12}=\tau_{8,12}\rho_{8,8}+\tau_{14,12}\rho_{14,14}-(\tau_{12,8}+\tau_{12,14})\rho_{12,12} +\sum_{j=4,10,11}(\Gamma_{j,12}\rho_{j,j}-\Gamma_{12,j}\rho_{12,12}) \\
&\dot{\rho}_{13,13}=\tau_{11,13}\rho_{11,11}-\tau_{13,11}\rho_{13,13}+\sum_{j=5,14,15}(\Gamma_{j,13}\rho_{j,j}-\Gamma_{13,j}\rho_{13,13}) \\
&\dot{\rho}_{14,14}=\tau_{12,14}\rho_{12,12}-\tau_{14,12}\rho_{14,14}+\sum_{j=6,13,16}(\Gamma_{j,14}\rho_{j,j}-\Gamma_{14,j}\rho_{14,14}) \\
&\dot{\rho}_{15,15}=\sum_{j=7,13,16}(\Gamma_{j,15}\rho_{j,j}-\Gamma_{15,j}\rho_{15,15}) \\
&\dot{\rho}_{16,16}=\sum_{j=8,14,15}(\Gamma_{j,16}\rho_{j,j}-\Gamma_{16,j}\rho_{16,16}) 
\label{eq:set_last}
\end{align}
In the above set of Eqns.~\eqref{eq:set_first}-\eqref{eq:set_last}, $\Gamma_{x,y}$ is related to the rate of state transition $\ket{x}\rightarrow \ket{y}$ due to system-to-reservoir tunnelling, and is dependent on the system-to-reservoir coupling as well as the probability of occupancy of the reservoirs. $\tau_{x,y}$, on the other hand, is related to the rate of state transition $\ket{x}\rightarrow \ket{y}$ due to inter-dot tunnelling and is to be derived from the density matrix equations. The various parameters $\Gamma_{x,y}$  can be given by the following equations, assuming statistical quasi-equilibrium at the reservoirs. 
\begin{align}
& \Gamma_{1,2}=\Gamma_{3,4}=\Gamma_{9,10}=\Gamma_{11,12}=\gamma_c f_G(\varepsilon_G) 
\label{eq:gamma_first} \\
& \Gamma_{5,6}=\Gamma_{7,8}=\Gamma_{13,14}=\Gamma_{15,16}=\gamma_cf_G(\varepsilon_G+U_m) \\
& \Gamma_{2,1}=\Gamma_{4,3}=\Gamma_{10,9}=\Gamma_{12,11}=\gamma_c\{1-f_G(\varepsilon_G)\} \\
& \Gamma_{6,5}=\Gamma_{8,7}=\Gamma_{14,13}=\Gamma_{16,15}=\gamma_c\{1-f_G(\varepsilon_G+U_m)\}\\
& \Gamma_{1,9}=\Gamma_{2,10}=\Gamma_{3,11}=\Gamma_{4,12}=\Gamma_{5,13}=\Gamma_{6,14}=\Gamma_{7,15}=\Gamma_{8,16}=\gamma_cf_L(\varepsilon_L) \\
& \Gamma_{9,1}=\Gamma_{10,2}=\Gamma_{11,3}=\Gamma_{12,4}=\Gamma_{13,5}=\Gamma_{14,6}=\Gamma_{15,7}=\Gamma_{16,8}=\gamma_c\{1-f_L(\varepsilon_L)\} \\
& \Gamma_{1,3}=\Gamma_{2,4}=\Gamma_{5,7}=\Gamma_{6,8}=\Gamma_{9,11}=\Gamma_{10,12}=\Gamma_{13,15}=\Gamma_{14,16}=\gamma_cf_R(\varepsilon_R) \\
& \Gamma_{3,1}=\Gamma_{4,2}=\Gamma_{7,5}=\Gamma_{8,6}=\Gamma_{11,9}=\Gamma_{12,10}=\Gamma_{15,13}=\Gamma_{16,14}=\gamma_c\{1-f_R(\varepsilon_R)\} 
\label{eq:gamma_last}
\end{align}
In Eqns.~\eqref{eq:set_first}-\eqref{eq:set_last}, for the system under consideration, the various values of $\tau_{x,y}$ were derived as:
\begin{equation}
\tau_{x,y}=\frac{t^2\Upsilon_{x,y}}{(\epsilon_x-\epsilon_y)^2+\left(\frac{\Upsilon_{x,y}}{2}\right)^2},
\end{equation}
where the different values of $\Upsilon_{x,y}$ can be expressed in terms of the following equations:
\begin{align}
&\Upsilon_{5,3}=\Upsilon_{3,5}=\sum_{j=6,7,13}\Gamma_{5,j}+\sum_{j=1,4,
11}\Gamma_{3,j}=\gamma_c\left\{1+2f_L(\varepsilon_L)+f_G(\varepsilon_G)+f_G(\varepsilon_G+U_m)\right\}
\label{eq:upsilon_first} \\
&\Upsilon_{4,6}=\Upsilon_{6,4}=\sum_{j=2,3,12}\Gamma_{4,j}+\sum_{j=5,8,14}\Gamma_{6,j}=\gamma_c\left\{3+2f_L(\varepsilon_L)-f_G(\varepsilon_G)-f_G(\varepsilon_G+U_m)\right\}\\
&\Upsilon_{5,9}=\Upsilon_{9,5}=\sum_{j=6,7,13}\Gamma_{5,j}+\sum_{j=1,10,11}\Gamma_{9,j}=\gamma_c\left\{1+2f_R(\varepsilon_R)+f_G(\varepsilon_G)+f_G(\varepsilon_G+U_m)\right\}\\
&\Upsilon_{6,10}=\Upsilon_{10,6}=\sum_{j=5,8,14}\Gamma_{6,j}+\sum_{j=2,9,12}\Gamma_{10,j}=\gamma_c\left\{3+2f_R(\varepsilon_R)-f_G(\varepsilon_G)-f_G(\varepsilon_G+U_m)\right\}\\
&\Upsilon_{7,11}=\Upsilon_{11,7}=\sum_{j=5,8,15}\Gamma_{7,j}+\sum_{j=3,9,12}\Gamma_{11,j}=\gamma_c\left\{3-2f_R(\varepsilon_R)+f_G(\varepsilon_G)+f_G(\varepsilon_G+U_m)\right\}\\
&\Upsilon_{8,12}=\Upsilon_{12,8}=\sum_{j=6,7,16}\Gamma_{8,j}+\sum_{j=4,10,11}\Gamma_{12,j}=\gamma_c\left\{5-2f_R(\varepsilon_R)-f_G(\varepsilon_G)-f_G(\varepsilon_G+U_m)\right\}\\
&\Upsilon_{11,13}=\Upsilon_{13,11}=\sum_{j=3,9,12}\Gamma_{11,j}+\sum_{j=5,14,15}\Gamma_{13,j}=\gamma_c\left\{3-2f_L(\varepsilon_L)+f_G(\varepsilon_G)+f_G(\varepsilon_G+U_m)\right\}\\
&\Upsilon_{12,14}=\Upsilon_{14,12}=\sum_{j=4,10,11}\Gamma_{12,j}+\sum_{j=6,13,16}\Gamma_{14,j}=\gamma_c\left\{5-2f_L       (\varepsilon_L)-f_G(\varepsilon_G)-f_G(\varepsilon_G+U_m)\right\}
\label{eq:upsilon_last}
\end{align}

From Eqn.~\eqref{eq:upsilon_first}-\eqref{eq:upsilon_last} and \eqref{eq:gamma_first}-\eqref{eq:gamma_last}, the different expressions for the inter-dot tunnelling rates were derived as:
\begin{align}
& \tau_{5,3}=\tau_{3,5}=\frac{t^2\Upsilon_{5,3}}{(\epsilon_M-\epsilon_R)^2+\left(\frac{\Upsilon_{5,3}}{2}\right)^2} \\
& \tau_{4,6}=\tau_{6,4}=\frac{t^2\Upsilon_{6,4}}{(\epsilon_M+U_m-\epsilon_R)^2+\left(\frac{\Upsilon_{6,4}}{2}\right)^2} \\
&\tau_{5,9}=\tau_{9,5}=\frac{t^2\Upsilon_{5,9}}{(\epsilon_M-\epsilon_L)^2+\left(\frac{\Upsilon_{5,9}}{2}\right)^2} \\
&\tau_{6,10}=\tau_{10,6}=\frac{t^2\Upsilon_{6,10}}{(\epsilon_M+U_m-\epsilon_L)^2+\left(\frac{\Upsilon_{6,10}}{2}\right)^2} \\
&\tau_{7,11}=\tau_{11,7}=\frac{t^2\Upsilon_{7,11}}{(\epsilon_M-\epsilon_L)^2+\left(\frac{\Upsilon_{7,11}}{2}\right)^2} \\
&\tau_{8,12}=\tau_{12,8}=\frac{t^2\Upsilon_{12,8}}{(\epsilon_M+U_m-\epsilon_L)^2+\left(\frac{\Upsilon_{8,12}}{2}\right)^2} \\
&\tau_{11,13}=\tau_{13,11}=\frac{t^2\Upsilon_{11,13}}{(\epsilon_M-\epsilon_R)^2+\left(\frac{\Upsilon_{11,13}}{2}\right)^2} \\
&\tau_{12,14}=\tau_{14,12}=\frac{t^2\Upsilon_{12,14}}{(\epsilon_M+U_m-\epsilon_R)^2+\left(\frac{\Upsilon_{12,14}}{2}\right)^2} \label{eq:tau_last}
\end{align}
The set of  Eqns. \eqref{eq:set_first}-\eqref{eq:set_last} form a set of linear equations. Once the different parameters are calculated using Eqns. \eqref{eq:gamma_first}-\eqref{eq:tau_last}, the diagonal elements of the density matrix, given by Eqns. \eqref{eq:set_first}-\eqref{eq:set_last}, can be solved using matrix inversion method, under the condition $\sum_{j=1}^{16}\rho_{j,j}=1$. Once the diagonal elements of the density matrix are computed, 
the current $I_{L(R)}$ through the left (right) reservoirs into the system can be given as:
\begin{align}
I_{L(R)}=\frac{q^2}{h}\gamma_c \left\{P^{\varsigma_{L(R)}}_0 f_{L(R)}(\varepsilon_{L(R)})-P^{\varsigma_{L(R)}}_1 \{1-f_{L(R)}(\varepsilon_{L(R)})\}\right\},
\label{eq:current}
\end{align}
where $P^{\varsigma_{L(R)}}_j$ denotes the probability that the ground state of the dot $S_{L(R)}$ contains $j$ electrons. $P^{\varsigma_{L(R)}}_j$ can be calculated by taking the partial trace over the density matrix elements as:
\begin{align}
\nonumber
&P^{\varsigma_{L}}_0=\sum_{j=1}^8 \rho_{j,j} \\
\nonumber
&P^{\varsigma_{L}}_1=\sum_{j=9}^{16} \rho_{j,j} \\
\nonumber
&P^{\varsigma_{R}}_0=\sum_{j=0}^3 (\rho_{4j+1,4j+1}+\rho_{4j+2,4j+2}) \\
&P^{\varsigma_{R}}_1=\sum_{j=0}^3 (\rho_{4j+3,4j+3}+\rho_{4j+4,4j+4}) 
\label{eq:system_prob}
\end{align}

\section{Derivation of $T_G^{opt}$ (Fig.~\ref{fig:Fig_7}) }\label{app_b}

	Here we provide a detailed explanation of  Eqn. \eqref{eq:1}. 
	To find the  remote reservoir temperature $T_G$ at which maximum sensitivity is achieved for a given  value of Coulomb-coupling energy $U_m$, we  need to analyze the maxima of the functions $\frac{df_G(\varepsilon_G)}{dT_G}|_{\varepsilon_G=0}$ and $\frac{d}{dT_G}f_G(\varepsilon_G+U_m)|_{\varepsilon_G=0}$.
	\begin{equation}
	f_G(\epsilon)=\left(1+exp\left\{\frac{\epsilon}{kT_{G}}\right\}\right)^{-1},
	\end{equation} 
	In the above equation, we assume $\mu_0=0$. In case of the set-up under consideration, we need to take into account the impact of both $\frac{df_G(\varepsilon_G)}{dT_G}|_{\varepsilon_G=0}$ and $\frac{d}{dT_G}f_G(\varepsilon_G+U_m)|_{\varepsilon_G=0}$ on the system performance.  However, $\frac{df_G(\varepsilon_G)}{dT_G}|_{\varepsilon_G=0}=0$ and thus the system sensitivity magnitude is maximized when $\frac{d}{dT_G}f_G(\varepsilon_G+U_m)|_{\varepsilon_G=0}$ is maximum. 
	The derivative of the function $f_G(\varepsilon_{G}+U_m)$ with respect to $T_G$ can be expressed as follows:
	\begin{equation}
	\frac{d}{dT_G}f_G(\varepsilon_G+U_m)=\left(\frac{\varepsilon_G+U_m}{kT_G^{2}}\right)\times exp\left(\frac{\varepsilon_G+U_m}{kT_{G}}\right) \times \left(1+exp\left\{\frac{\varepsilon_G+U_m}{kT_{G}}\right\}\right)^{-2},
	\end{equation}
	\begin{figure*}[!htb]
		\centering
		\includegraphics[width=1.02\textwidth]{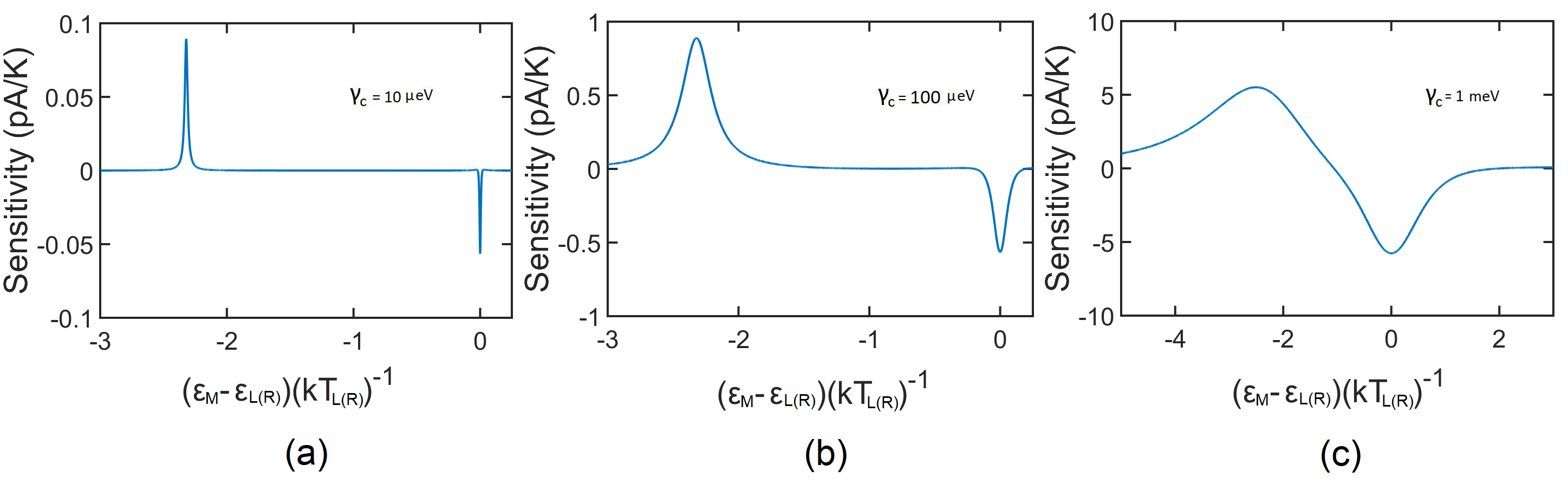}
		\caption{Variation in  system-I sensitivity  with the ground state misalignment  $\varepsilon_M-\varepsilon_{L(R)}$ for (a) $\gamma_c=10\upmu$eV (b) $\gamma_c=100\upmu$eV (c) $\gamma_c=1$meV. The Coulomb-coupling energy is assumed to be $U_m=2$meV$~(\approx 2.3209\frac{kT}{q})$. The bias V is chosen as $V=0.0129$V$~(\approx 15\frac{kT}{q})$ and the temperature of the reservoirs $L$, $R$ and $G$ are taken as $T_L$=$T_R$=$T_G=10$K. The ground states $\varepsilon_L$, $\varepsilon_R$ and $\varepsilon_G$ are aligned with $\mu_0$, that is, $\varepsilon_L=\varepsilon_R=\varepsilon_G=\mu_{0}$}. 
		\label{fig:Fig_10}
	\end{figure*}
	For further simplification,  assuming $\left(\frac{\varepsilon_G+U_m}{k}\right)=c$, and $\frac{1}{T_G}=x$, we can express  $\frac{d}{dT_G}f_G(\varepsilon_G+U_m)$ as:
	\begin{equation}
	\frac{d}{dT_G}f_G(\varepsilon_G+U_m)=cx^2 \times exp(cx) \times \left(1+exp(cx)\right)^{-2},
	\end{equation}
		On differentiating the above equation (to obtain the point of maxima), and after some algebraic manipulations, we obtain an expression as:
	\begin{equation}
	\frac{d^2}{dT_G^2}f_G(\varepsilon_G+U_m)=0 \Rightarrow (1+exp(cx)) \times (cx+2) = 2 \times cx \times exp(cx)
	\Rightarrow \frac{cx+2}{cx-2}=exp(cx)
	\end{equation}
	After some algebraic manipulations, we obtain the expression as:
	\begin{equation}
	\frac{exp\left(\frac{\varepsilon_G+U_m}{kT_G^{opt}}\right)+1}{exp\left(\frac{\varepsilon_G+U_m}{kT_G^{opt}}\right)-1} = 0.5 \times \left(\frac{\varepsilon_{G}+U_m}{kT_G^{opt}}\right)
	\end{equation}
We note that the above equation is satisfied for  $\left(\frac{U_m}{kT_G^{opt}}\right)=2.399$. Hence, the optimal remote reservoir  temperature $T_G$ corresponding to maximum magnitude of sensitivity is obtained is given by:
	\begin{equation}
	T_{G}^{opt}=(2.399)^{-1}\frac{\varepsilon_G+U_m}{  k}
	\end{equation}
	For $\varepsilon_{G}=0$, the value of $T_G^{opt}$ is hence given by $T_G^{opt}=(2.399)^{-1}\frac{U_m}{  k}$
\section{Optimal bias voltage for maximum sensitivity at different $U_m$ for system-II. }\label{app_c}
\begin{table}[ht]
		
		\centering
		\begin{tabular}{|c | c| }
			\hline
			Coulomb-coupling energy ($U_m$)   &    Optimal bias $V$  $\Big($in $\frac{kT_{L(R)}}{q}\Big)$ \\
			\hline
			$1 ~$meV & $2.974$   \\ 
			$2~ $meV & $3.674$   \\
			$3 ~$meV & $4.723$  \\
			$4~ $meV & $5.773$  \\
			$5 ~$meV & $6.823$  \\

			\hline
		\end{tabular}
	\caption{Table for different values of $U_m$ and the corresponding optimal voltage bias for which maximum sensitivity of system-II was achieved.}
		\label{table1}
	\end{table}
\end{widetext}
  \section{Realistic implementation of the quadruple quantum dot thermometer}
  	\begin{figure*}[!htb]
  	\centering
  	\includegraphics[width=1.04\textwidth]{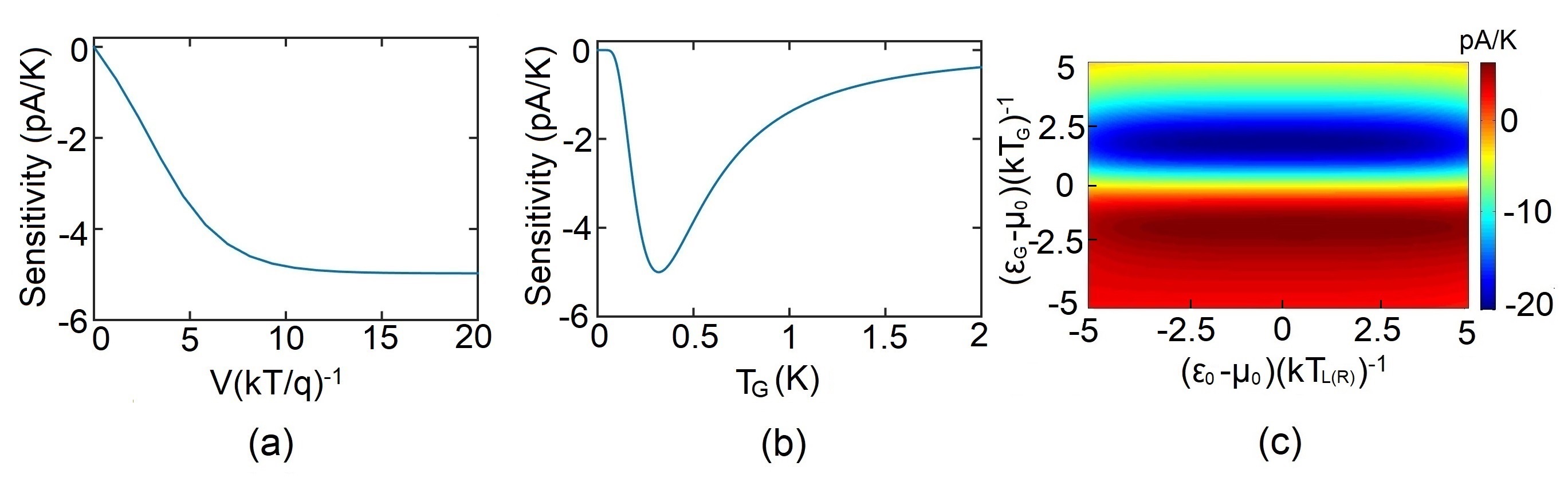}
  	\caption{Performance of the thermometer (system-I) for the set of parameters obtained from a recent experiment by Thierschmann, \emph{et. al.}  \cite{heatengine13}. The chosen parameter values are chosen from Ref. \cite{heatengine13} as $\gamma_c=30\upmu$eV, $U_m=70\upmu$eV, $T_L=T_R=300$mK. (a) Variation in thermometer sensitivity with applied bias $V$, for $T_L=T_R=T_G=300$mK (b) Variation in thermometer sensitivity with $T_G$ (c) Variation in sensitivity with quantum dot ground states. The ground states of $S_L,~S_M$ and $S_R$ are assumed to be aligned with each other as $\varepsilon_0=\varepsilon_{L}=\varepsilon_{M}=\varepsilon_{R}$. }
  	\label{fig:Fig_11}
  \end{figure*} 
  \indent Despite a higher sensitivity and robustness against voltage noise or fluctuation in the average current path temperature, the quadruple quantum dot thermometer demands extreme fine-tuning to align the ground states of $S_L,~S_M$ and $S_R$. Such fine-tuning of energy levels is extremely difficult in practice. A practical method to align the quantum dot ground states may be the fabrication of identical size quantum dots. In addition, gate-defined ground states tuning also offers an alternative method. Besides,  one may also ensure non-zero temperature sensitivity with slightly misaligned ground states by enhancing the system-to-reservoir coupling, which broadens the quantum dot ground states and provides electron transmission over a wider energy range. Fig.~\ref{fig:Fig_10} demonstrates the change in temperature sensitivity regime as the system-to-reservoir coupling is increased from $10\upmu$eV to $1$meV. We note that with an increase in the system-to-reservoir coupling, the thermometer becomes fairly tolerant to some misalignment in ground states and operates over a wider regime of $\varepsilon_{M}-\varepsilon_{L(R)}$.  \color{black}
\section{Investigation of the quadruple quantum dot thermometer for realistic  parameters obtained from a recent experiment by  Thierschmann, \emph{et. al.} \cite{heatengine13}}\label{app_d}
The purpose of the present manuscript was to elaborate that the quadruple quantum dot thermometer can operate over a wide range of parameters, such as $U_m,~T_G,~T_{L(R)}$. The parameters' range, used in this paper, are within the scope of experimental feasibility. In addition, coming to the  quantum dot charging energy, systems with mutual Coulomb-coupling energy ranging as high as $0.3$meV to $2$meV and self-charging energy (due to self-capacitance) of $7$meV and more have already been fabricated  \cite{cap_coup_1,cap_coup_2,cap_coup3,extra}. We believe that with further progress in solid-state nano fabrication technology, a stronger Coulomb-coupling between spatially separated quantum dots is not too far behind and can be achieved in the near future. Since  experimental parameters can vary drastically depending on the quantum dot fabrication technology, in this section we demonstrate the performance of the thermometer for a particular set of  parameters obtained from a recent experiment by Thierschmann, \emph{et. al.} \cite{heatengine13}. In the recent experimental work, Thierschmann, \emph{et. al.}, have fabricated and analyzed a non-local heat engine based on Coulomb-coupled quantum dots \cite{heatengine13}. The system parameters were calculated from experimental data in the Supplementary Section of Ref. \cite{heatengine13}, and were found to lie in the range,
\begin{align}
& \gamma_c=20 \mathrm{\upmu eV} \sim 50\mathrm{\upmu eV} \nonumber \\
& U_m \approx 70\mathrm{\upmu eV} \nonumber \\
& T_G,T_L,T_R  =\mathrm{230mK} \sim \mathrm{380mK}
\end{align}  
\indent To demonstrate the performance of the set-up with a particular set of experimental parameters, we assume $\gamma_c=30\upmu$eV, $U_m=70\upmu$eV, $T_G=T_L=T_R=300$mK, which are directly taken from the range of experimental parameters calculated in the Supplementary Section of Ref.~\cite{heatengine13}. Fig.~\ref{fig:Fig_11} demonstrates the performance of the quadruple quantum dot thermometer with the above set of experimental parameters. We note that with the above set of parameters, the thermometer operates fruitfully, albeit over a different regime. Thus the proposed design can pave the way towards practical implementation of non-local cryogenic thermometers.

 \bibliography{apssamp}
\end{document}